# Achieving High Temporal Resolution in Single-Molecule Fluorescence Techniques using Plasmonic Nanoantennas

Sunny Tiwari,[1] Prithu Roy,[1] Jean-Benoît Claude,[1] Jérôme Wenger[1],*

[1] Aix Marseille Univ, CNRS, Centrale Marseille, Institut Fresnel, AMUTech, 13013 Marseille, France

* Corresponding author: jerome.wenger@fresnel.fr

**Abstract:**

Single-molecule fluorescence techniques are essential for investigating the molecular mechanisms in biological processes. However, achieving sub-millisecond temporal resolution to monitor fast molecular dynamics remains a significant challenge. The fluorescence brightness is the key parameter that generally defines the temporal resolution for these techniques. Conventional microscopes and standard fluorescent emitters fall short in achieving the high brightness required for sub-millisecond monitoring. Plasmonic nanoantennas have been proposed as a solution, but despite huge fluorescence enhancement have been obtained with these structures, the brightness generally remains below 1 million photons/s/molecule. Therefore, the improvement of temporal resolution has been overlooked. In this article, we present a method for achieving high temporal resolution in single-molecule fluorescence techniques using plasmonic nanoantennas, specifically optical horn antennas. We demonstrate about 90% collection efficiency of the total emitted light, reaching a high fluorescence brightness of 2 million photons/s/molecule in the saturation regime. This enables observations of single molecules with microsecond binning time and fast fluorescence correlation spectroscopy (FCS) measurements. This work expands the applications of plasmonic antennas and zero-mode waveguides in the fluorescence saturation regime towards brighter single-molecule signal, faster temporal resolutions and improved detection rates to advance fluorescence sensing, DNA sequencing and dynamic studies of molecular interactions.

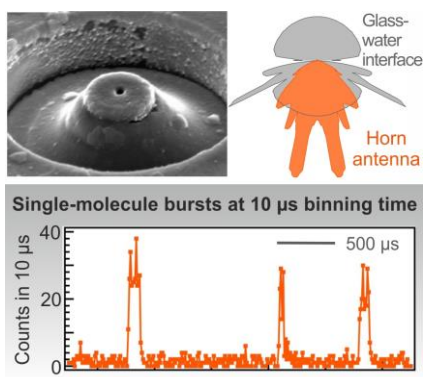

**Keywords:** optical nanostructures, plasmonics, optical antennas, single molecule fluorescence, light-matter interaction

Figure for Table of Contents

## 1. Introduction

Single-molecule fluorescence techniques are essential to investigate and understand the molecular mechanisms behind biological processes.[1–5] A broad class of single-molecule spectroscopy methods has been developed, including fluorescence time trace analysis,[6] fluorescence correlation spectroscopy (FCS),[7] Förster resonance energy transfer (FRET),[8,9] single particle tracking (SPT),[10] or multiparameter fluorescence detection (MFD).[11,12] With these techniques, watching single molecule dynamics on timescales from milliseconds to minutes is particularly common. However, achieving a sub-millisecond temporal resolution to monitor fast molecular dynamics remains highly challenging.[2,13–19]

The fluorescence brightness (detected number of photons per second and per molecule) is the key parameter generally defining the temporal resolution for the different single-molecule fluorescence techniques. With a typical single-molecule fluorescence brightness of 50,000 photons/s/molecule, it takes 1 ms to collect 50 photons and reliably detect a fluorescence signal. However, if the recording time is set to 100 µs, then only 5 photons are detected on average within this time interval, leading to a large shot noise contribution and a poor signal-to-noise ratio. To obtain 50 photons within 10 µs, then the fluorescence brightness must reach 5 million photons/s/molecule. Unfortunately, these high fluorescence brightness and their associated fast temporal resolutions remain out of reach of the conventional confocal microscopes and standard fluorescent emitters.[20] The large size mismatch between a single molecule and the wavelength of light implies that the phenomenon of diffraction ultimately limits the light-matter interaction in conventional microscopes and their ability to collect the energy radiated by the single molecule.[21,22]

Plasmonic nanoantennas have been introduced to overcome the diffraction limit in light-matter interactions.[22,23] Impressive fluorescence enhancement factors above 1000-fold have been achieved using bowtie,[24] nanorod,[25,26] nanoparticle assemblies,[27] DNA-origami dimer,[28–32] DNA-templated dimer,[33–35] or antenna-in-box antennas.[36,37] These large enhancements are generally achieved using low quantum yield emitters and by avoiding saturation at moderate excitation intensities, in order to maximize the nanoantenna enhancement. While different studies on DNA origami nanoantennas have reported high fluorescence brightness above 1 million photons per second for a single molecule,[28,29,32] the majority of research in this field focuses on maximizing the fluorescence enhancement factor rather than the net detected fluorescence brightness.

Improving the temporal resolution of single-molecule fluorescence techniques requires detecting several photons within a microsecond, which in turn implies a fluorescence brightness above the million photons/s/molecule. The fluorescence enhancement is not the primary interest, what matters most here is the absolute brightness or photon count rate per molecule. To reach the high detection rates required, the excitation power needs to be increased so as to reach the fluorescence saturation regime, which results from the finite lifetime of the excited state and the buildup of triplet or dark states.[20] In the saturation regime, the fluorescence brightness per emitter is no longer proportional to the excitation intensity and the quantum yield, but depends on the product of the collection efficiency times the radiative decay rate constant.[38–40,32] This slightly changes the usual design rules for plasmonic nanoantennas to enhance fluorescence, and calls for a specific maximization of the collection efficiency and the radiative rate enhancement. Despite the crucial role that fluorescence saturation plays in determining the maximum fluorescence brightness, only a few studies have explicitly investigated the antenna-emitter coupling at saturation.[38–40,32] Up to 100-fold increase in the maximum photon count rate at saturation could be achieved using gold nanorods,[39,40] yet photophysical processes, such as the formation of dark states and photoisomerization, were found to still limit the fluorescence brightness at saturation in the nanoantenna.[32] While the aforementioned studies were instrumental in pioneering the field, the nanoantenna influence in improving the temporal resolution in biological applications still remains largely unexplored.

Here, we specifically target the use of plasmonic antennas to maximize the fluorescence brightness per molecule in the saturation regime and simultaneously improve the temporal resolution dynamics of single-molecule techniques. As the collection efficiency and radiative rate enhancement are the two important parameters for the brightness at saturation, we choose to work with optical horn antennas, which feature a conical horn microreflector combined with a central nanoaperture milled into an aluminum film and are the optical analogue of microwave horn antennas (Fig. 1a-c).[41] The rationale behind this design is two fold: (i) the nanoaperture of 120 nm diameter serves as an attoliter analysis chamber locally enhancing the fluorescence emission rate,[38,42] while (ii) the conical reflector directs the fluorescence light towards the microscope objective and improves the total collection efficiency.[43–45] The nanoaperture brings the advantages of the zero-mode waveguides (ZMW) to improve the detection of fluorescence molecules in solution with higher brightness and 1000-fold higher concentration than standard diffraction-limited microscopes.[46–52] The conical reflector allows to collect the so-called supercritical or forbidden light emitted at angles above the critical angle for total internal reflection at a glass-water interface.[21,53] Because its working principle does not depend on interference effects unlike Yagi Uda or Bull's eye antennas for directivity control,[54,55] the horn antenna operates over a broad spectral range largely covering the full fluorescence emission spectrum. Thanks to the central aperture, the horn antenna remains fully compatible with the detection of fluorescent molecules diffusing in solution as the emitter does not need to be embedded into the antenna material contrarily to dielectric[56–59] or patch antenna designs.[60–63] Our group has recently developed UV horn antennas designed to improve the autofluorescence detection sensitivity of label-free proteins in the ultraviolet.[43,44] However, this antenna design has never been used in the visible spectral range and never in the context of fluorescence at saturation.

Here, we demonstrate close to 90% collection efficiency of the total emitted light from a single dipole enabling to reach a high fluorescence brightness of 2 million photons/s/molecule in the saturation regime. We discuss how the dark state buildup is the main factor limiting the maximum photon count rate and show that the local temperature increase plays a significant role in the saturation condition. The large fluorescence brightness achieved with the horn antenna enables observing single molecules within individual diffusion bursts with a binning time as short as 10 μs. It also allows fast FCS measurements with integration times of only 200 ms to monitor high affinity molecular interaction dynamics as we demonstrate for streptavidin-biotin association at micromolar concentrations. This work further expands the applications of plasmonic antennas and ZMWs towards higher fluorescence count rates and faster temporal resolutions to improve nanopore sensing[64–67] DNA sequencing,[68–71] as well as dynamic studies of molecular interactions,[72–79] protein conformations,[18,80] and biomembrane organization.[81–86]

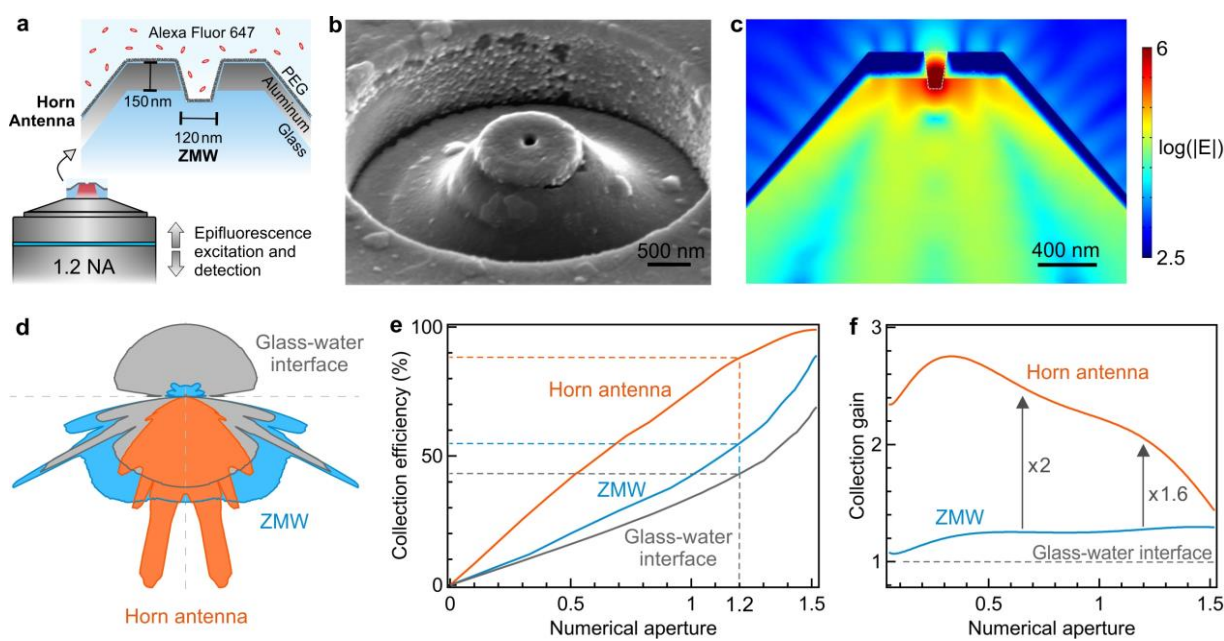

**Figure 1.** Optical horn antennas to achieve 90% fluorescence collection efficiency from single emitters. (a) Scheme of the experimental configuration. (b) Scanning electron microscope image of a horn antenna with a 120 nm diameter ZMW nanoaperture milled in the centre. (c) Numerical simulations of the electric field intensity radiated by a single dipole positioned in the centre of the nanoaperture with vertical orientation and 670 nm emission wavelength. (d) Calculated three-dimensional radiation patterns from dipolar sources (averaged over x, y, and z orientations) placed in the horn antenna, ZMW, and 5 nm above a glass-water interface. (e) Collection efficiency as a function of the numerical

aperture used for the microscope objective. (f) Gain in collection efficiency for the horn antenna and the ZMW respective to the glass-water interface as a function of the numerical aperture used for the microscope objective. The numbers near the arrows indicate the extra gains brought by the horn antenna respective to the ZMW for the two numerical apertures used in this work.

## 2. Results and Discussion

Our main experiments are performed with Alexa Fluor 647 dyes (excitation 633 nm, emission peak at 670 nm), which are among the dyes with the highest radiative rate constants (Supporting Information Fig. S1) and are thus good candidates to maximize the photon count rate in saturation conditions. Figure 1a shows a scheme of the experimental configuration. A single horn antenna fabricated using focused ion beam (FIB) (Fig. 1b and S2) is excited using a focused 633 nm CW laser beam in an epifluorescence confocal microscope (See Experimental Section for details). The center of the horn antenna comprises a nanoaperture of 120 nm diameter chosen to maximize the fluorescence brightness enhancement based on our previous works on ZMWs.[38,87] The 43° angle for the conical reflector is chosen based on direct experimental characterization of different microreflectors in the visible (Fig. S3) together with the consideration of our available data in the UV.[43] Throughout this work, we will compare the full horn antenna (ZMW + microreflector) with the bare ZMW without microreflector. This provides an important reference to assess the influence of the microreflector in improving the collection efficiency. The numerical simulations reported in Fig. S4 show that the microreflector does not modify the local intensity inside the nanoaperture: both the horn antenna and the bare ZMW show very similar electromagnetic intensity profiles inside the 120 nm aperture. This is expected as the excitation beam is focused by a 1.2 numerical aperture (NA) objective to a diffraction-limited spot of 700 nm diameter smaller than the top plateau of the horn antenna.

The reflective wall of the horn antenna redirects most of the emission in a narrow range of angles towards the collection objective as shown in the numerical simulations in Fig. 1c,d and S5. For comparison, the radiation patterns for a bare ZMW and a glass-water interface are also plotted on Fig. 1d. To quantify the beam steering ability of the horn antenna, we calculate the collection efficiency as a function of the numerical aperture used in the experiment for three different configurations: horn antenna, ZMW, and 5 nm above a glass-water interface (Fig. 1e). For a 1.2 NA microscope objective as used in our experiments, 43% of the total emitted light is collected just above the glass-water interface in the confocal configuration. This collection efficiency increases to 55% for a bare ZMW while the horn antenna achieves an impressive collection of 90% of the total radiated power in all 3 space directions. Our simulations predict a collection gain of 1.6-fold for the horn antenna as compared to the ZMW for a 1.2 NA objective. This gain is even higher for lower NA objectives (Fig. 1f), as a consequence of the horn antenna directing the majority of the emission into a narrow ±20° cone (Fig. 1d).

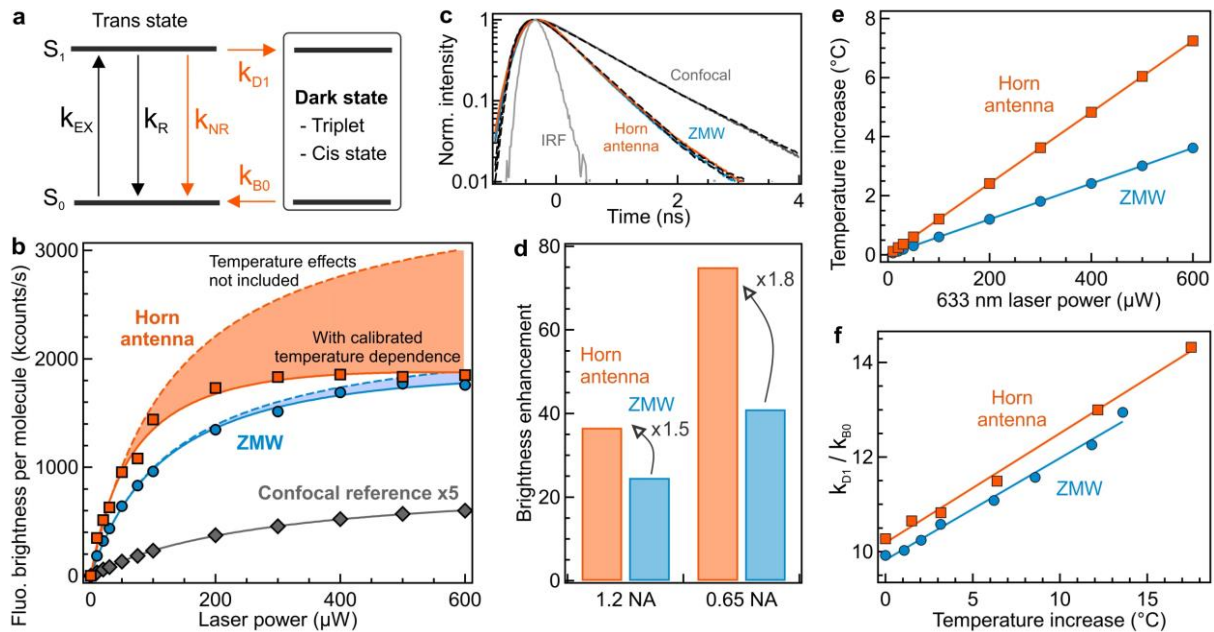

**Figure 2.** Fluorescence in the saturation regime: reaching high fluorescence brightness above 1 million counts per second and per molecule with the horn antennas and the influence of the local temperature. (a) Notations for the photophysical rate constants (see text for details). (b) Fluorescence brightness per molecule as a function of the laser power for the horn antenna, the ZMW and the confocal reference. The curves are numerical fits using the model Eq. (1) with and without considering the temperature effects on the transition rates. (c) Normalised fluorescence lifetime decay curves showing an equal lifetime reduction for the horn antenna and the ZMW as compared to the confocal reference. IRF stands for the instrument response function. (d) Fluorescence brightness enhancement for the horn antenna and the ZMW probed with two different objective numerical apertures at a laser power of 50 µW. (e) Temperature increase in the horn antenna and ZMW as a function of the 633 nm laser power. The markers are experimental data deduced from infrared measurements (Fig. S11), the lines are numerical fits. (f) Ratio of the decay rate constants $k_{D1}/k_{B0}$ leading to the dark state build-up as a function of the temperature increase. The markers are experimental data deduced from FCS measurements (Fig. S12), the lines are numerical fits.

Figure 2 summarizes our experimental characterizations of the fluorescence enhancement and the fluorescence brightness in the linear and saturation regimes. As a cyanine dye, Alexa Fluor 647 undergoes trans-cis photoisomerization leading to a non-emissive state in the microsecond time scale.[88,89] Since our FCS experiments cannot distinguish between photoisomerization and triplet blinking, we decide to represent the photophysics of Alexa Fluor 647 with the model pictured in Fig. 2a where $k_{EX}$, $k_R$ the radiative, $k_{NR}$ are respectively the excitation, radiative and nonradiative rate constants of the fluorescent trans conformation, $k_{D1}$ is the transition rate constant from S$_1$ to the dark state and $k_{B0}$ the recovery rate constant from the dark state back to the ground state S$_0$ in trans conformation.[88,89] $k_{D1}$ and $k_{B0}$ include contributions from both trans-cis photoisomerization and triplet transitions, providing a model with the least number of parameters to conclusively describe our data. With these notations, the fluorescence brightness (count rate per molecule $CRM$), is given by[20,90]

$$CRM = \kappa \frac{k_R}{k_R + k_{NR}} \frac{k_{EX}}{1 + \frac{k_{EX}}{k_R + k_{NR}}\left(1 + \frac{k_{D1}}{k_{B0}}\right)} \quad (1)$$

where $\kappa$ denotes the collection efficiency. In the quantum efficiency term $k_R/(k_R + k_{NR})$, we have neglected the transition rate constant $k_{D1}$ as it is three orders of magnitude smaller than the radiative and nonradiative rate constants.[88] For high excitation rates in the saturation condition $k_{EX} \gg k_R + k_{NR}$, the molecular brightness becomes

$$CRM_{saturation} = \kappa \frac{k_R}{1 + \frac{k_{D1}}{k_{B0}}} \quad (2)$$

As discussed in the introduction, this equation reminds that the fluorescence brightness in the saturation condition depends on the collection efficiency $\kappa$, the radiative rate constant $k_R$ and the build-up $k_{D1}/k_{B0}$ of the dark state.

To measure the fluorescence brightness in our different experiments, we perform fluorescence correlation spectroscopy (FCS) which allows to record the average fluorescence intensity and the average number of detected molecules for each experimental run (see Experimental Section and Fig. S6-S9 for details).[7] Figure 2b is our main experimental result: while in the confocal configuration the molecular brightness $CRM$ saturates at values below 120 kcounts/s, the horn antenna brings it close to 2 million photons per second and per molecule (Mcounts/s/molecule), providing a 16-fold improvement over the net maximum detected count rate per molecule as compared to the diffraction-limited case. Surprisingly, the horn antenna gain as compared to the ZMW is significant for low and moderate excitations powers (below 100 µW) but tends to decrease as the saturation conditions are fully reached at powers above 400 µW. This feature is unexpected as the collection efficiency gain should be constant for all different excitation powers (Eq. (1-2)). To explain the different observations in Fig. 2b, we need to address separately the experimental gain in collection efficiency and the physical origins of the fluorescence saturation in the horn antenna.

First, we ensure that the horn antenna and the ZMW modify similarly the local density of optical states (LDOS): fluorescence lifetime measurements taken in similar conditions to the FCS experiments indicate a similar fluorescence lifetime for Alexa 647 in the horn antenna as in the ZMW, with a lifetime reduction from 1.0 ns in the confocal configuration to 0.57 ns in the horn antenna and the ZMW (Fig. 2c). This result agrees with the numerical simulations in Fig. S4 indicating similar electric field distributions inside the nanoaperture regardless of the presence of the microreflector, which affects only the directivity of the emitted light. This demonstrates that at low excitation powers below saturation, the decay rate constants $k_{EX}$, $k_R$, $k_{NR}$ are similar between the horn antenna and the ZMW.

Comparing the fluorescence brightness obtained with the horn antenna and the bare ZMW below saturation, Eq. (1) indicates that their brightness difference amounts to the gain in the collection efficiency $\kappa$. Experimentally, we find a 1.5-fold collection gain for the horn antenna as compared to the bare ZMW below the fluorescence saturation and for a 1.2 NA objective (Fig. 2d). This result agrees very well with the numerical simulations in Fig. 1f where a 1.6-fold gain was predicted. The same set of experiments is reproduced with an objective of lower 0.65 NA (Fig. S10). In this case, we find a larger collection gain of 1.8-fold with the horn antenna respective to the ZMW (Fig. 2d), again in good agreement with the numerical simulations in Fig. 1f. Altogether, these results demonstrate the beam steering ability of the horn antenna which contributes to further enhance the detected fluorescence brightness as compared to the ZMW.

If the horn antenna was only redirecting the fluorescence emitted at high angles towards the microscope objective, then the fluorescence brightness in the horn antenna should always be 1.5-fold higher than the ZMW brightness, following the orange dashed line in Fig. 2d and reaching values around 3 Mphotons/s/molecule. However, our experimental data clearly deviate from this trend for excitation powers in the saturation regime above 100 µW. The FCS results of the number of molecules and diffusion time do not indicate any major difference between the horn antenna and the ZMW (Fig. S7), ruling out any extra photobleaching that would occur predominantly in the horn antenna as compared to the ZMW (the fast diffusion times below 50 µs and the continuous illumination avoid the conditions leading to strong photobleaching).[90]

In the presence of metal and with local excitation intensities exceeding several hundreds of µW/µm², the local temperature increase due to the absorption of the excitation light into the metal (Joule heating) cannot be neglected. A first hint about the occurrence of this process is provided by the decrease of the FCS diffusion time with increased excitation powers (Fig. S7d) as a result of the lower viscosity of water at elevated temperatures[91] and the generation of thermos-osmotic flows accelerating the apparent diffusion.[92] We measure the local temperature inside the horn antenna and the bare ZMW using the fluorescence lifetime protocol introduced in [93] which is summarized in the Supporting Information section S11. Figure 2e shows our experimental values of the temperature increase in the horn antenna and the ZMW as a function of the 633 nm excitation power. We confirm that the experimental results for the ZMW are consistent with the numerical simulations performed in [94]. For the same input laser power, the temperature increase in the horn antenna is about twice that in the ZMW, reaching non-negligible values of +5°C at 400 µW. We explain the higher temperature elevation in the horn antenna by the fact that the aluminum layer only extends to a few µm and is discontinuous to the rest of the metal film, as a consequence of the whole horn antenna being milled into the substrate (Fig. 1b). For the ZMW milled into the planar metal film, the heat is more efficiently evacuated as the ZMW is directly connected to the 1-inch aluminum layer covering the whole substrate.[95]

The local temperature increase affects the fluorescence photophysics in two major ways. First, it increases the nonradiative decay rate $k_{NR}$ and reduces the fluorescence lifetime.[88,93] However, at fluorescence saturation, Eq. (2) shows that $k_{NR}$ does not play a major role (the influence in the quantum yield and the saturation intensity tend to balance each other). Second, the temperature also affects the trans-cis photoisomerization process, [88,89] modifying the dark state transition rate constants $k_{D1}$ and $k_{B0}$. We probe this effect using supplementary FCS measurements discussed in the Supporting Information section S12. Both rate constants increase with the temperature (Fig. S12),[88] the most important result being that the dark state build-up $k_{D1}/k_{B0}$ follows a linear dependence with the local temperature (Fig. 2f). The validity of our results is confirmed by the fact that both the horn antenna and the ZMW lead to similar dependences of $k_{D1}/k_{B0}$ with the local temperature (Fig. 2f). As the excitation power increases to reach fluorescence saturation, the local temperature simultaneously rises, leading to an increase of the dark state build-up $k_{D1}/k_{B0}$ quantity which turns to decrease the fluorescence brightness at saturation (denominator in Eq. (2)). This effect is more pronounced for the horn antenna than the ZMW because of the twice higher temperature increase in the case of the horn antenna (Fig. 2e). As a consequence of the temperature affecting $k_{D1}/k_{B0}$, the brightness observed experimentally with the horn antenna (markers in Fig. 2b) significantly deviates from the behaviour expected in the absence of any temperature elevation (dashed orange line). Using the experimentally measured values of the local temperature as well as the temperature dependence of $k_{NR}$, $k_{D1}$ and $k_{B0}$, we get a nice interpolation of the horn antenna brightness data (Fig. 2b solid orange curve) with no free parameter as compared to the ZMW fit.

Altogether, the results in Fig. 2 show that high fluorescence brightness per molecule around 2 Mphotons/s can be reached with the horn antenna thanks to its improved collection efficiency in agreement with numerical simulations. The brightness at saturation is currently limited by the dark state build-up $k_{D1}/k_{B0}$ which depends on the local temperature elevation. This phenomenon is not limited to Alexa 647 dyes, our experiments on Alexa 546 molecules report similar observations (Fig. S13), while it was noted that in the case of DNA origami nanoantennas, the fluorescence brightness in the nanoantenna was still limited by the underlying photophysical processes, such as formation of dim states and photoisomerization.[32] For advanced applications, sapphire substrates may dissipate the extra heat more efficiently than glass,[94] yet we have found that gallium-FIB milling of horn antennas into sapphire substrates was producing a too high luminescent background so that more advanced nanofabrication techniques (helium-FIB, reactive ion etching of the luminescent contaminants…) will be needed for such applications.

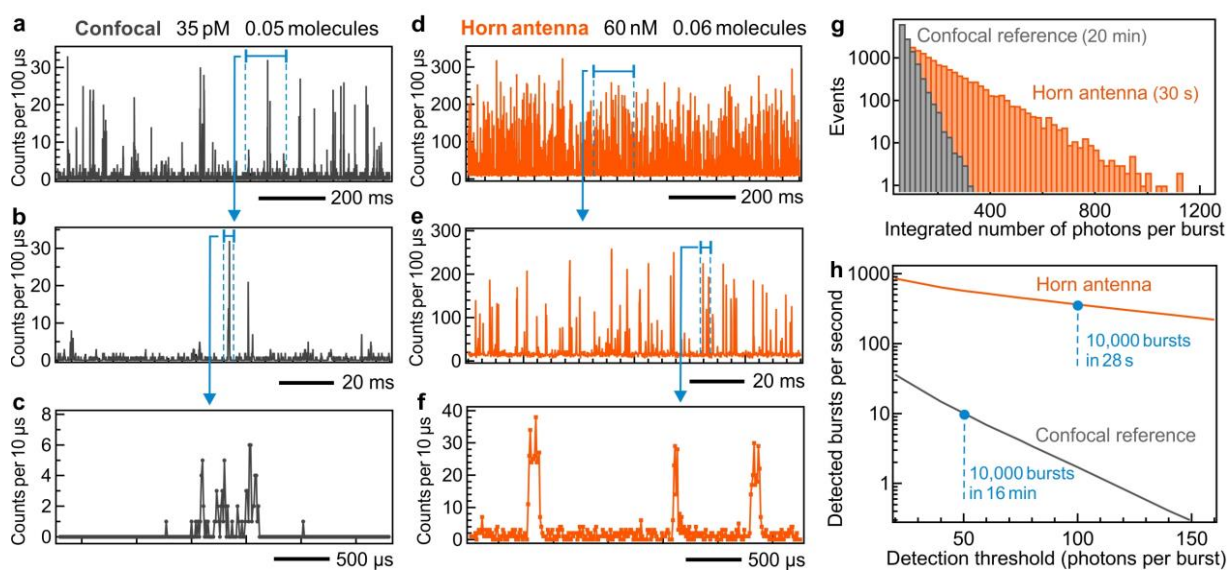

**Figure 3.** Watching single molecule bursts at 10 microsecond resolution thanks to the high fluorescence brightness from horn antennas. (a-c) Fluorescence time traces for the confocal reference showing typical bursts from diffusing single molecules with a binning time of 100 µs (a,b) and 10 µs (c). (d-f) Same as (a-c) for the horn antenna. The Alexa 647 concentrations are chosen so as to have comparable average numbers of molecules detected by FCS in both configurations (Supporting Information Fig. S14). The laser power is 200 µW. (g) Histograms of the total numbers of photons per burst. The integration time in the confocal reference is 20 min while for the horn antenna it is 30 s. (h) Number of detected bursts per second as a function of the detection threshold for the configurations corresponding to (a,d).

Next, we use the high fluorescence brightness per molecule achieved with the horn antenna to monitor single molecule dynamics at a fast µs binning time. Figure 3a-f shows the fluorescence intensity time trace of single Alexa 647 molecules diffusing at low concentration. For both the confocal (Fig. 3a-c) and horn antenna configuration (Fig. 3d-f), we select the Alexa 647 concentration so as to have a similar 0.05 average number of molecules for both cases (this feature is controlled by FCS, see Fig. S14). This low average number of molecules ensures that the bursts stem from single diffusing molecules. For the confocal and the horn antenna, the concentrations are 35 pM and 60 nM respectively.

In the confocal configuration, binning the time trace with a 10 µs window poorly resolves even the brightest bursts (Fig. 3c). This is a consequence of the moderate confocal brightness of 75 kcounts/s/molecule, which means that in a 10 µs time period less than 1 photon is detected on average. Moving to the horn antenna, the brightness becomes 1,7 Mcounts/s/molecule, so that an average of 17 photons can be detected with 10 µs, which enables directly monitoring the fluorescence time trace at microsecond resolution (Fig. 3f). This feature illustrates how the high fluorescence brightness directly translates into an improved temporal resolution to monitor single molecule dynamics. The high fluorescence brightness with the horn antenna is also clearly visible in the number of detected photons per time bin (vertical axis in Fig. 3a-f) as well as in the integrated total number of photons per burst (histogram in Fig. 3g and scatter plot in Fig. S15). For completeness, the fluorescence time trace at 10 µs binning time for the ZMW is shown on Fig. S16, while the comparative analysis on single molecule bursts in Fig. S15 confirms the superior performance of the horn antenna. Besides, the fluorescence brightness derived from the single-molecule burst analysis in Fig. S15a,c confirm the FCS results in Fig.2b.

Another striking feature while comparing the times traces for the confocal setup and the horn antenna (Fig. 3b and 3d) is that a >20-fold higher number of intense fluorescence bursts is seen with the horn antenna as compared to the confocal reference. This happens although the concentrations were chosen to have a similar FCS average number of molecules detected in both cases (Fig. S14). A detailed analysis using the burst search module of *PAM* software (see Experimental section for details) confirms the higher number of detected bursts per second for the horn antenna (Fig. 3h and S16). With the horn antenna, 10,000 bursts of 100 photons each are collected within only 28 seconds, while about 16 minutes are needed in the confocal reference to achieve a similar total number of bursts with a twice lower threshold of 50 photons per burst (Fig. 3g). The data in Fig. S16 compare the detection performance of the horn antenna and the ZMW. For similar experimental conditions and detection threshold, the horn antenna detects 30% more single-molecule bursts per second owing to its improved fluorescence brightness.

The higher detection rate of intense bursts with the horn antenna is a consequence of two features: (i) because of the reduced detection volume in the horn antenna, the diffusion time and the burst duration are reduced by about 4-fold so that a 4-fold higher number of diffusion events can occur during one second, and (ii) because of the enhanced brightness with the horn antenna, the bursts are better resolved above the detection threshold. For the horn antenna with an average number of 0.06 molecules and a burst duration of 120 µs, 10 detection events are expected within a 20 ms window, which corresponds to the data in Fig. 3d. However, for the confocal reference with 0.05 molecules and an average burst duration of 500 µs, 2 detection events are expected per 20 ms, or equivalently 12 single-molecule diffusion events for the whole 125 ms duration of Fig. 3b. This is clearly not what is observed while using a standard detection threshold of 50 photons per burst in confocal mode (a common practice in this field[9]). The conclusion is that typically >80% detection events are missed in the confocal configuration. In Fig. 3b, this corresponds to the small fluorescence fluctuations with only a few photons per 100 µs, which are not noise but undetected single molecule diffusion events. The enhanced brightness achieved with the horn antenna efficiently solves this issue, making all single-molecule diffusion events easily detectable against the background.

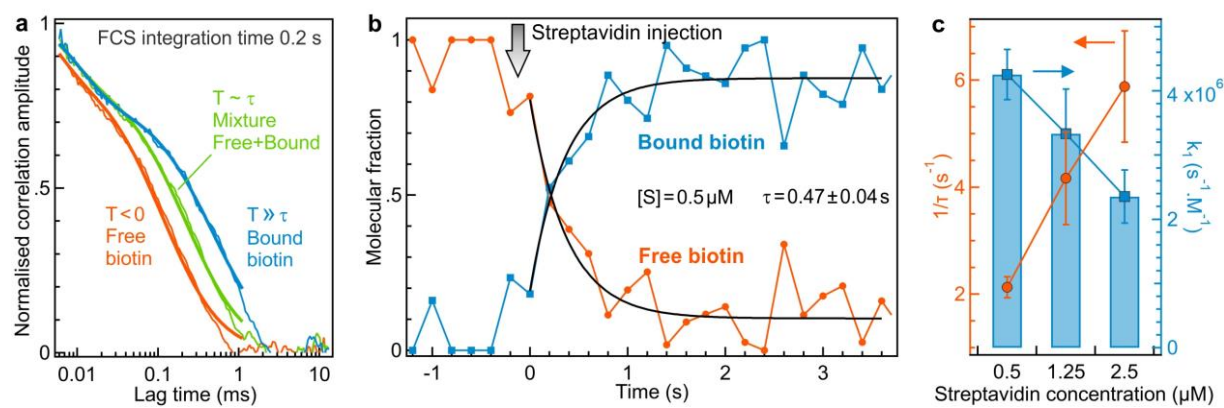

**Figure 4.** High fluorescence brightness in the horn antenna enables fast FCS measurements to monitor high affinity association rates of streptavidin and biotin. (a) Amplitude-normalized FCS correlation curves of Atto 643-biotin mixed with the streptavidin solution at T = 0. The integration time to compute each FCS trace is 200 ms. The 3.8-fold shift in the diffusion time indicates binding of biotin to streptavidin. (b) Temporal evolution of the free and bound molecular fractions of biotin after adding a 0.5 µM solution of streptavidin to a 0.5 µM biotin solution. Each data point results from a 200 ms FCS measurement. $\tau$ is the characteristic time of the exponential evolutions (see Supporting Information section S17). (c) Evolution of the inverse of the association time $\tau$ (orange markers, left axis) and the association rate constant $k_1$ (blue bars, right axis) as a function of the streptavidin concentration.

Lastly, we use the high count rate per molecule achieved using the horn antenna to quantify the association rate constant of biotin with streptavidin with FCS at micromolar concentrations. Biotin binding to streptavidin has major applications in biotechnology and is a prime example of a biochemical reaction with a very high affinity. The binding rate constant of streptavidin and biotin was found to be in a range between 3.0 × 10$^6$ to 4.5 × 10$^7$ M$^{-1}$.s$^{-1}$.[96] At a 1 µM concentration, the characteristic association time becomes only between 20 and 330 ms. This time is too short to be probed using confocal FCS, which typically requires over 5 s integration time.[7,97] However, the high brightness with the horn antenna allows to drastically reduce the FCS integration time, down to 200 ms as shown here (Fig. 4a).

Our experiments use biotins labelled with Atto-643 at 0.5 µM concentration which are mixed with label-free streptavidin at concentrations from 0.5 to 2.5 µM. These conditions ensure that a single biotin binds to a streptavidin, and that cooperativity between the 4 different binding sites can be discarded here. Upon binding to streptavidin, the Atto643-biotin diffusion time gradually evolves from 190 µs (free biotin) to 730 µs (fully bound biotin) (Fig. 4a). This 3.8-fold increase in the diffusion time is consistent with the change of molecular weight from 1.26 kDa (free Atto643-biotin) to 67 kDa (Atto643-biotin-streptavidin complex) as $\sqrt[3]{67/1.26} \sim 3.8$.[97] We use FCS fits with two species to determine the respective ratios of free and bound biotin (see Experimental section for details). The high fluorescence brightness enables FCS with 0.2 s integration time, enough to monitor the streptavidin-biotin association even at micromolar concentrations. The minimum FCS integration time reachable with the horn antenna is no longer dictated only by the signal to noise ratio, but rather by the necessity in FCS to have an integration time at least 200-fold greater than the diffusion time in order to avoid artifacts.[98]

The temporal evolution of the free and bound fractions are plotted in Fig. 4b and S17 for different streptavidin concentrations in the micromolar range. Upon binding to streptavidin, the fraction of free biotin declines exponentially with a characteristic time $\tau = 1/(k_1[S])$, where $k_1$ is the association rate constant and $[S]$ the streptavidin concentration (see a general discussion in the Supporting Information section S17). Fitting the free and bound fractions to extract the characteristic time $\tau$ (Fig. 4c, left axis) then allows to calculate the association rate constant $k_1$ (Fig. 4c, right axis). We find $k_1$ values between 2.4 × 10$^6$ to 4.3 × 10$^6$ M$^{-1}$.s$^{-1}$ in close agreement to the reported values using a FRET readout in droplet-based microfluidics.[96] Using the dissociation rate constant $k_{-1}$ of 2 × 10$^{-6}$ s$^{-1}$ determined in [99,100], the dissociation constant $K_d = k_{-1}/k_1$ can be estimated on the order of 10$^{-14}$ M, in agreement with the commonly reported value.[101] For the aim of this paper, the data in Fig. 4 demonstrate that the enhanced brightness in the horn antenna opens the possibility to perform fast FCS measurements on the challenging high-affinity association kinetics at micromolar concentrations.

## 3. Conclusions

We have demonstrated the use of plasmonic nanoantennas to achieve high temporal resolutions in single-molecule fluorescence techniques. By using optical horn antennas, we were able to achieve close to 90% collection efficiency of the total emitted light from a single dipole, resulting in a high fluorescence brightness of 2 million photons/s/molecule in the saturation regime. We also discussed the role of dark state buildup and local temperature increase in limiting the maximum photon count rate in the saturation condition. The high fluorescence brightness achieved with the horn antenna enables the observation of single molecules with a binning time as short as 10 µs and fast FCS measurements at 200 ms integration time. This opens up new possibilities for monitoring high-affinity molecular interactions at micromolar concentrations as demonstrated in our example of streptavidin-biotin association. One key benefit of the horn antenna is that it greatly improves the detection efficiency compared to confocal setups. Specifically, we observed a > 20-fold higher number of fluorescence bursts with the horn antenna as compared to the confocal reference, making it an effective solution for missed detection events that are common in confocal configurations. On the other hand, the horn antenna and ZMW approaches necessitate concentrations of several tens of nanomolar, which could present a challenge for certain protein applications in cases where synthesis has low production yields and struggles to attain concentrations exceeding a few nanomolar. Furthermore, the horn antenna concept can be extended to improve single-photon sources and non-linear light emitting devices. While impressive results have been achieved using planar dielectric or plasmonic patch antennas, [56–63] these designs embed the emitter into the dielectric material, which is not ideal for biological applications. The horn antenna presented in this study is designed to keep the fluorescent molecule in a water environment and at room temperature, improving temporal resolution in single-molecule fluorescence analysis. However, the orientation of the molecular emitter cannot be controlled and the design does not provide as strong fluorescence enhancement as other plasmonic designs. Our work expands the applications of plasmonic antennas and ZMWs towards higher fluorescence count rates and faster temporal resolutions, which can improve single molecule fluorescence sensing, DNA sequencing and dynamic studies of molecular interactions, protein conformations, and biomembrane organization.

## 4. Experimental section

*Nanofabrication.* The horn antennas and ZMW nanoapertures are fabricated by focused ion beam (FIB, Fig. S1).[43] A conical structure is first milled by FIB (FEI DB235 Strata, 30 kV acceleration voltage, 300 pA ion current). A 100 nm thick aluminum layer is then deposited by electron-beam evaporation (Bühler Syrus Pro 710) to make the cone walls reflective. A 120 nm diameter aperture is milled by FIB (10 pA current) in the center of the microreflector to realize the horn antenna. For the ZMW, the nanoaperture is milled elsewhere on the sample with the same FIB conditions to provide a direct comparison with the horn antenna. Lastly, the aluminum samples are protected by a 12 nm-thick silica layer deposited with plasma-enhanced chemical vapor protection (Oxford Instruments PlasmaPro NGP80).[102]

*Numerical simulations.* The electric field profile was simulated using finite element method with the wave optics module of COMSOL Multiphysics v5.5. Nanoapertures of diameter 120 nm in both the horn antenna and ZMW are excited from the bottom using a Gaussian source of wavelength 633 nm. The refractive index of the glass substrate and water (in the aperture and the upper medium) are set as 1.52 and 1.33 respectively, whereas, the refractive index of aluminum is taken from ref.[103] A user defined mesh of size ranging from 0.3 nm to 10 nm is set to create the mesh in a tetrahedral geometry. To suppress the reflection from the boundaries, scattering boundary conditions is used. To calculate the Fourier plane images, the calculated near-field profile in COMSOL is transformed to the far-field using RETOP an open-source MATLAB toolbox.[104] The horn antenna or ZMW is excited using an individual oscillating dipole in x, y, and z directions. The wavelength of the oscillations is set at 670 nm which is near the emission maxima of the Alexa 647 dyes used in the experiments. The Fourier plane images are calculated by incoherently adding the radiation pattern of the emission from each individual x, y, and z oriented dipoles. The Fourier plane image in the case of horn antenna shows confinement in the emission angles as compared to a broader emission in the case of ZMW.

*Sample preparation.* Alexa 647 dyes purchased from ThermoFisher are used as received and are diluted in Phosphate-Buffered Saline (PBS) solution. Prior to use, the horn antenna and ZMW surface are passivated using polyethylene glycol (PEG) silane (Nanocs PG1-SL-1K) of 1000 Da molar mass following the protocol in [105]. Prior to passivation, the samples are cleaned with water, followed by cleaning with ethanol (97%) and rinsing with isopropyl alcohol. After this, the samples are exposed under UV Ozone for 10 minutes to clean any organic impurities. This cleaning process with water, ethanol, and IPA followed by UV Ozone, is repeated thrice to completely clean the surface. Finally, the samples are put in air plasma cleaner for 5 minutes. After the plasma cleaning, the samples are immediately covered with 1 mM PEG solution prepared in absolute ethanol with 1% acetic acid. The chamber is blown with argon gas and left at room temperature for 4 hours. The samples are rinsed with ethanol and dried using nitrogen. To store the passivated samples, a 1% Tween20 solution in absolute ethanol is used.

Biotin molecules tagged with Atto 643 dyes are purchased from ATTO-TEC. Streptavidin extracted from *Streptomyces avidinii* is purchased from Sigma-Aldrich and its supernatant is separated and stored at -20°C after centrifuging at 142000 g for 12 minutes using Airfuge 20 psi Ultracentrifuge. For Figure 2, the concentrations of Alexa 647 dyes in PBS solution are 600 nM for the horn antenna and the ZMW, and 1 nM for the confocal reference. All the molecular concentrations are measured using Tecan Spark 10 M Spectrofluorometer.

*Experimental setup.* The FCS measurements are performed using a custom-built confocal microscope (Nikon Ti-U Eclipse) equipped with both continuous 633 nm HeNe laser and 635 nm pulsed laser (LDH series laser diode, PicoQuant, pulse duration ~ 50 ps). A multiband dichroic mirror (ZT 405/488/561/640rpc, Chroma) is used to reflect both lasers towards the microscope. The molecules are excited using either 63x, 1.2 NA water immersion (Zeiss C-Apochromat) or 40x 0.65 Air (Olympus PlanN) objective lenses and the emission is collected in the epifluorescence configuration. To efficiently reject the backscattered laser, the emission is first passed through the same multiband dichroic mirror (ZT 405/488/561/640rpc, Chroma) and then through two emission filters (ZET405/488/565/640mv2 and ET655, Chroma). The collected signal was focused onto the single-photon avalanche photodiodes (APD) (Perkin Elmer SPCM AQR 13) after passing through either 50 or 100 μm pinholes to spatially filter the molecular fluorescence. A supplementary emission filter (FF01-676/37, Semrock) was placed in front of the APD to block the entering of stray light in the APD, and further reject the Rayleigh scattered light. The photon counting events are detected in a time-tagged time-resolved mode using a fast time-correlated single photon counting module (HydraHarp 400, Picoquant). Our avalanche photodiode is given for a dead time of 50 ns. At the highest count rate of 2 Mphotons/s detected here, the correction factor for the APD dead time is 1.11. We have decided to neglect this correction here in order to ease the comparison with other experiments.

*FCS analysis.* All the FCS curves are fitted using the following equation where we use a three dimensional Brownian diffusion model with an additional term to include the blinking effects.[7,87,105]

$$G(\tau) = \rho \left[ 1 + \frac{T_{ds}}{1-T_{ds}} \exp\left(-\frac{\tau}{\tau_{ds}}\right) \right] \left(1 + \frac{\tau}{\tau_d}\right)^{-1} \left(1 + \frac{1}{s^2}\frac{\tau}{\tau_d}\right)^{-0.5} \quad (3)$$

Where $G(\tau)$ is the correlation amplitude, $\rho$ is the correlation amplitude, $T_{ds}$ is the fraction of molecules going in the dark state, $\tau_{ds}$ is the blinking time of the dark state, $\tau_d$ is the mean diffusion time, and $\kappa$ corresponds to the aspect ratio of the axial to the transversal dimension of the excitation source in the solution. For the confocal case, we used a value of $s$=5, whereas for the optical horn antennas, based on our past results for the ZMW aperture, we used a value of $s$=1 which fits well the experimental correlation amplitude. To extract the fitting parameters such as the number of molecules and the diffusion time, for the confocal case, we fit the experimental data from 3 μs to 10 ms lag time, whereas for the ZMW and horn antenna, the data was fitted from 3 µs to 0.1 ms, as the fitting model starts to divert from the experimental data at longer lag times. From the fitting values of the correlation amplitude, $\rho$, the measured background, $B$, on the horn antenna, and ZMW, and the total detected intensity, $F$, we extract the average number of molecules $N$ and the brightness or count rate per molecule $CRM$ as:

$$N = \frac{1}{\rho}\left(1 - \frac{B}{F}\right)^2 \quad (4)$$

$$CRM = \frac{1}{N}(F - B) \quad (5)$$

In FCS, the average number of molecules $N$ is defined as the mean number of detected fluorescent molecules in the observation volume averaged over the duration of the experiment.[7]

*Fluorescence burst analysis.* We use the *Burst Analysis* module of the PIE Analysis with MATLAB (PAM).[106] The burst detection is performed using the *All Photon Burst Search* function by putting the threshold photon counts, the time window of the bursts, and the total photon counts per burst. Different values of the threshold for the photons in a burst, minimum photons per time window, and burst time window are taken for the horn antenna, ZMW, and confocal reference due to the difference in the diffusion time and the background signal in all the three cases. For the confocal configuration, each peak is considered as a single molecule burst having at least 50 photons in the burst and a minimum of 2 photons per burst time window of 100 µs. For the horn antenna (ZMW) the minimum photons per burst is 100 (75) with a minimum of 12 (10) photons per burst time window of 50 (50) µs. The value of the burst time window is taken near the extracted diffusion time from the FCS analysis for the horn antenna, ZMW, and the confocal reference. For figure 3h, the value of the burst detection threshold (minimum photons per burst) is changed by keeping the other parameters constant for all the three configurations.

**Supporting Information**

Radiative rate constants of fluorescent dyes, Fabrication of the horn antenna, influence of the microreflector cone angle, near field enhancement in the aperture, beaming using horn antenna, FCS time traces and correlation functions, FCS numbers of molecules and diffusion times, Background intensities, statistical dispersion between different structures, collection gain with 0.65 NA, temperature measurements, Dark state transition rate constants of Alexa Fluor 647 with temperature, fluorescence brightness of Alexa 546, FCS data corresponding to the burst analysis, statistics of brightness and photon count per burst, single molecule bursts in ZMW, biotin-streptavidin association rate calculation, measurement of biotin-streptavidin association rate.

**Acknowledgements**

The authors thank Don Lamb for stimulating discussions. This project has received funding from the European Research Council (ERC) under the European Union's Horizon 2020 research and innovation programme (grant agreement No 723241).

**Conflict of Interest**

The authors declare no conflict of interest.

**Data Availability Statement**

The data that support the findings of this study data are available from the corresponding author upon request.

**References**

[1]  T. Ha, P. Tinnefeld, *Annu. Rev. Phys. Chem.* **2012**, *63*, 595.

[2]  B. Schuler, H. Hofmann, *Curr. Opin. Struct. Biol.* **2013**, *23*, 36.

[3]  M. Orrit, *Angew. Chem. Int. Ed.* **2015**, *54*, 8004.

[4]  C. Joo, H. Balci, Y. Ishitsuka, C. Buranachai, T. Ha, *Annu. Rev. Biochem.* **2008**, *77*, 51.

[5]  S. Dey, M. Dolci, P. Zijlstra, *ACS Phys. Chem. Au* **2023**, DOI: 10.1021/acsphyschemau.2c00061.

[6]  H. P. Lu, L. Xun, X. S. Xie, *Science* **1998**, *282*, 1877.

[7]  T. Wohland, S. Maiti, R. Machan, *An Introduction to Fluorescence Correlation Spectroscopy*, IOP Publishing Ltd, **2020**.

[8]  E. Lerner, T. Cordes, A. Ingargiola, Y. Alhadid, S. Chung, X. Michalet, S. Weiss, *Science* **2018**, *359*, eaan1133.

[9]  B. Hellenkamp, S. Schmid, O. Doroshenko, O. Opanasyuk, R. Kühnemuth, S. R. Adariani, B. Ambrose, M. Aznauryan, A. Barth, V. Birkedal, M. E. Bowen, H. Chen, T. Cordes, T. Eilert, C. Fijen, C. Gebhardt, M. Götz, G. Gouridis, E. Gratton, T. Ha, P. Hao, C. A. Hanke, A. Hartmann, J. Hendrix, L. L. Hildebrandt, V. Hirschfeld, J. Hohlbein, B. Hua, C. G. Hübner, E. Kallis, A. N. Kapanidis, J.-Y. Kim, G. Krainer, D. C. Lamb, N. K. Lee, E. A. Lemke, B. Levesque, M. Levitus, J. J. McCann, N.

Naredi-Rainer, D. Nettels, T. Ngo, R. Qiu, N. C. Robb, C. Röcker, H. Sanabria, M. Schlierf, T. Schröder, B. Schuler, H. Seidel, L. Streit, J. Thurn, P. Tinnefeld, S. Tyagi, N. Vandenberk, A. M. Vera, K. R. Weninger, B. Wünsch, I. S. Yanez-Orozco, J. Michaelis, C. A. M. Seidel, T. D. Craggs, T. Hugel, *Nat. Methods* **2018**, *15*, 669.

[10] A. Kusumi, T. A. Tsunoyama, K. M. Hirosawa, R. S. Kasai, T. K. Fujiwara, *Nat. Chem. Biol.* **2014**, *10*, 524.

[11] C. Eggeling, S. Berger, L. Brand, J. R. Fries, J. Schaffer, A. Volkmer, C. A. M. Seidel, *J. Biotechnol.* **2001**, *86*, 163.

[12] J. Widengren, V. Kudryavtsev, M. Antonik, S. Berger, M. Gerken, C. A. M. Seidel, *Anal. Chem.* **2006**, *78*, 2039.

[13] L. A. Campos, J. Liu, X. Wang, R. Ramanathan, D. S. English, V. Muñoz, *Nat. Methods* **2011**, *8*, 143.

[14] T. Otosu, K. Ishii, T. Tahara, *Nat. Commun.* **2015**, *6*, 7685.

[15] S. Kilic, S. Felekyan, O. Doroshenko, I. Boichenko, M. Dimura, H. Vardanyan, L. C. Bryan, G. Arya, C. A. M. Seidel, B. Fierz, *Nat. Commun.* **2018**, *9*, 235.

[16] H. Sanabria, D. Rodnin, K. Hemmen, T.-O. Peulen, S. Felekyan, M. R. Fleissner, M. Dimura, F. Koberling, R. Kühnemuth, W. Hubbell, H. Gohlke, C. A. M. Seidel, *Nat. Commun.* **2020**, *11*, 1231.

[17] A.-M. Cao, R. B. Quast, F. Fatemi, P. Rondard, J.-P. Pin, E. Margeat, *Nat. Commun.* **2021**, *12*, 5426.

[18] M. F. Nüesch, M. T. Ivanović, J.-B. Claude, D. Nettels, R. B. Best, J. Wenger, B. Schuler, *J. Am. Chem. Soc.* **2022**, *144*, 52.

[19] D. E. Makarov, A. Berezhkovskii, G. Haran, E. Pollak, *J. Phys. Chem. B* **2022**, *126*, 7966.

[20] J. Enderlein, C. Zander, in *Single Mol. Detect. Solut.*, Wiley-Blackwell, **2003**, pp. 21–67.

[21] L. Novotny, B. Hecht, *Principles of Nano-Optics*, Cambridge University Press, **2012**.

[22] L. Novotny, N. van Hulst, *Nat. Photonics* **2011**, *5*, 83.

[23] A. F. Koenderink, *ACS Photonics* **2017**, *4*, 710.

[24] A. Kinkhabwala, Z. Yu, S. Fan, Y. Avlasevich, K. Müllen, W. E. Moerner, *Nat. Photonics* **2009**, *3*, 654.

[25] H. Yuan, S. Khatua, P. Zijlstra, M. Yorulmaz, M. Orrit, *Angew. Chem. Int. Ed.* **2013**, *52*, 1217.

[26] S. Khatua, P. M. R. Paulo, H. Yuan, A. Gupta, P. Zijlstra, M. Orrit, *ACS Nano* **2014**, *8*, 4440.

[27] D. Punj, R. Regmi, A. Devilez, R. Plauchu, S. B. Moparthi, B. Stout, N. Bonod, H. Rigneault, J. Wenger, *ACS Photonics* **2015**, *2*, 1099.

[28] G. P. Acuna, F. M. Möller, P. Holzmeister, S. Beater, B. Lalkens, P. Tinnefeld, *Science* **2012**, *338*, 506.

[29] A. Puchkova, C. Vietz, E. Pibiri, B. Wünsch, M. Sanz Paz, G. P. Acuna, P. Tinnefeld, *Nano Lett.* **2015**, *15*, 8354.

[30] K. Trofymchuk, V. Glembockyte, L. Grabenhorst, F. Steiner, C. Vietz, C. Close, M. Pfeiffer, L. Richter, M. L. Schütte, F. Selbach, R. Yaadav, J. Zähringer, Q. Wei, A. Ozcan, B. Lalkens, G. P. Acuna, P. Tinnefeld, *Nat. Commun.* **2021**, *12*, 950.

[31] K. Trofymchuk, K. Kołątaj, V. Glembockyte, F. Zhu, G. P. Acuna, T. Liedl, P. Tinnefeld, *ACS Nano* **2023**, *17*, 1327–1334.

[32] L. Grabenhorst, K. Trofymchuk, F. Steiner, V. Glembockyte, P. Tinnefeld, *Methods Appl. Fluoresc.* **2020**, *8*, 024003.

[33] M. P. Busson, B. Rolly, B. Stout, N. Bonod, S. Bidault, *Nat. Commun.* **2012**, *3*, 962.

[34] S. Bidault, A. Devilez, V. Maillard, L. Lermusiaux, J.-M. Guigner, N. Bonod, J. Wenger, *ACS Nano* **2016**, *10*, 4806.

[35] A. P. Francisco, D. Botequim, D. M. F. Prazeres, V. V. Serra, S. M. B. Costa, C. A. T. Laia, P. M. R. Paulo, *J. Phys. Chem. Lett.* **2019**, *10*, 1542.

[36] D. Punj, M. Mivelle, S. B. Moparthi, T. S. van Zanten, H. Rigneault, N. F. van Hulst, M. F. García-Parajó, J. Wenger, *Nat. Nanotechnol.* **2013**, *8*, 512.

[37] V. Flauraud, R. Regmi, P. M. Winkler, D. T. L. Alexander, H. Rigneault, N. F. van Hulst, M. F. García-Parajó, J. Wenger, J. Brugger, *Nano Lett.* **2017**, *17*, 1703.

[38] J. Wenger, D. Gérard, J. Dintinger, O. Mahboub, N. Bonod, E. Popov, T. W. Ebbesen, H. Rigneault, *Opt. Express* **2008**, *16*, 3008.

[39] Y. Wang, M. Horáček, P. Zijlstra, *J. Phys. Chem. Lett.* **2020**, *11*, 1962.

[40] E. Wientjes, J. Renger, R. Cogdell, N. F. van Hulst, *J. Phys. Chem. Lett.* **2016**, *7*, 1604.

[41] C. A. Balanis, *Antenna Theory: Analysis and Design*, John Wiley & Sons, **2005**.

[42] J. de Torres, P. Ghenuche, S. B. Moparthi, V. Grigoriev, J. Wenger, *ChemPhysChem* **2015**, *16*, 782.

[43] A. Barulin, P. Roy, J.-B. Claude, J. Wenger, *Nat. Commun.* **2022**, *13*, 1842.

[44] P. Roy, J.-B. Claude, S. Tiwari, A. Barulin, J. Wenger, *Nano Lett.* **2023**, *23*, 497–504.

[45] T. Grosjean, M. Mivelle, G. W. Burr, F. I. Baida, *Opt. Express* **2013**, *21*, 1762.

[46] M. J. Levene, J. Korlach, S. W. Turner, M. Foquet, H. G. Craighead, W. W. Webb, *Science* **2003**, *299*, 682.

[47] P. Zhu, H. G. Craighead, *Annu. Rev. Biophys.* **2012**, *41*, 269.

[48] P. Holzmeister, G. P. Acuna, D. Grohmann, P. Tinnefeld, *Chem. Soc. Rev.* **2014**, *43*, 1014.

[49] D. Punj, P. Ghenuche, S. B. Moparthi, J. de Torres, V. Grigoriev, H. Rigneault, J. Wenger, *Wiley Interdiscip. Rev. Nanomed. Nanobiotechnol.* **2014**, *6*, 268.

[50] M. S. Alam, F. Karim, C. Zhao, *Nanoscale* **2016**, *8*, 9480.

[51] N. Maccaferri, G. Barbillon, A. N. Koya, G. Lu, G. P. Acuna, D. Garoli, *Nanoscale Adv.* **2021**, *3*, 633.

[52] G. M. Crouch, D. Han, P. W. Bohn, *J. Phys. Appl. Phys.* **2018**, *51*, 193001(1.

[53] T. Ruckstuhl, J. Enderlein, S. Jung, S. Seeger, *Anal. Chem.* **2000**, *72*, 2117.

[54] A. G. Curto, G. Volpe, T. H. Taminiau, M. P. Kreuzer, R. Quidant, N. F. van Hulst, *Science* **2010**, *329*, 930.

[55] H. Aouani, O. Mahboub, N. Bonod, E. Devaux, E. Popov, H. Rigneault, T. W. Ebbesen, J. Wenger, *Nano Lett.* **2011**, *11*, 637.

[56] K. G. Lee, X. W. Chen, H. Eghlidi, P. Kukura, R. Lettow, A. Renn, V. Sandoghdar, S. Götzinger, *Nat. Photonics* **2011**, *5*, 166.

[57] X.-L. Chu, T. J. K. Brenner, X.-W. Chen, Y. Ghosh, J. A. Hollingsworth, V. Sandoghdar, S. Götzinger, *Optica* **2014**, *1*, 203.

[58] W. Li, L. Morales-Inostroza, W. Xu, P. Zhang, J. Renger, S. Götzinger, X.-W. Chen, *ACS Photonics* **2020**, *7*, 2474.

[59] M. Colautti, P. Lombardi, M. Trapuzzano, F. S. Piccioli, S. Pazzagli, B. Tiribilli, S. Nocentini, F. S. Cataliotti, D. S. Wiersma, C. Toninelli, *Adv. Quantum Technol.* **2020**, *3*, 2000004.

[60] C. Belacel, B. Habert, F. Bigourdan, F. Marquier, J.-P. Hugonin, S. Michaelis de Vasconcellos, X. Lafosse, L. Coolen, C. Schwob, C. Javaux, B. Dubertret, J.-J. Greffet, P. Senellart, A. Maitre, *Nano Lett.* **2013**, *13*, 1516.

[61] T. B. Hoang, G. M. Akselrod, M. H. Mikkelsen, *Nano Lett.* **2016**, *16*, 270.

[62] S. Checcucci, P. Lombardi, S. Rizvi, F. Sgrignuoli, N. Gruhler, F. B. Dieleman, F. S Cataliotti, W. H. Pernice, M. Agio, C. Toninelli, *Light Sci. Appl.* **2017**, *6*, e16245.

[63] S. Morozov, M. Gaio, S. A. Maier, R. Sapienza, *Nano Lett.* **2018**, *18*, 3060.

[64] O. N. Assad, T. Gilboa, J. Spitzberg, M. Juhasz, E. Weinhold, A. Meller, *Adv. Mater.* **2017**, *29*, 1605442.

[65] D. V. Verschueren, S. Pud, X. Shi, L. De Angelis, L. Kuipers, C. Dekker, *ACS Nano* **2019**, *13*, 61.

[66] N. Klughammer, C. Dekker, *Nanotechnology* **2021**, *32*, 18LT01.

[67] F. Farhangdoust, F. Cheng, W. Liang, Y. Liu, M. Wanunu, *Adv. Mater.* **n.d.**, *n/a*, 2108479.

[68] J. Eid, A. Fehr, J. Gray, K. Luong, J. Lyle, G. Otto, P. Peluso, D. Rank, P. Baybayan, B. Bettman, A. Bibillo, K. Bjornson, B. Chaudhuri, F. Christians, R. Cicero, S. Clark, R. Dalal, A. DeWinter, J. Dixon, M. Foquet, A. Gaertner, P. Hardenbol, C. Heiner, K. Hester, D. Holden, G. Kearns, X. Kong, R. Kuse, Y. Lacroix, S. Lin, P. Lundquist, C. Ma, P. Marks, M. Maxham, D. Murphy, I. Park, T. Pham, M. Phillips, J. Roy, R. Sebra, G. Shen, J. Sorenson, A. Tomaney, K. Travers, M. Trulson, J. Vieceli, J. Wegener, D. Wu, A. Yang, D. Zaccarin, P. Zhao, F. Zhong, J. Korlach, S. Turner, *Science* **2009**, *323*, 133.

[69] S. Uemura, C. E. Aitken, J. Korlach, B. A. Flusberg, S. W. Turner, J. D. Puglisi, *Nature* **2010**, *464*, 1012.

[70] J. Chen, R. V. Dalal, A. N. Petrov, A. Tsai, S. E. O'Leary, K. Chapin, J. Cheng, M. Ewan, P. L. Hsiung, P. Lundquist, S. W. Turner, D. R. Hsu, J. D. Puglisi, *Proc. Natl. Acad. Sci. U. S. A.* **2014**, *111*, 664.

[71] J. Larkin, R. Y. Henley, V. Jadhav, J. Korlach, M. Wanunu, *Nat. Nanotechnol.* **2017**, *12*, 1169.

[72] K. T. Samiee, M. Foquet, L. Guo, E. C. Cox, H. G. Craighead, *Biophys. J.* **2005**, *88*, 2145.

[73] T. Miyake, T. Tanii, H. Sonobe, R. Akahori, N. Shimamoto, T. Ueno, T. Funatsu, I. Ohdomari, *Anal. Chem.* **2008**, *80*, 6018.

[74] T. Sameshima, R. Iizuka, T. Ueno, J. Wada, M. Aoki, N. Shimamoto, I. Ohdomari, T. Tanii, T. Funatsu, *J. Biol. Chem.* **2010**, *285*, 23159.

[75] J. Zhao, S. P. Branagan, P. W. Bohn, *Appl. Spectrosc.* **2012**, *66*, 163.

[76] T. Sandén, R. Wyss, C. Santschi, G. Hassaïne, C. Deluz, O. J. F. Martin, S. Wennmalm, H. Vogel, *Nano Lett.* **2012**, *12*, 370.

[77] Y. Zhao, D. Chen, H. Yue, M. M. Spiering, C. Zhao, S. J. Benkovic, T. J. Huang, *Nano Lett.* **2014**, *14*, 1952.

[78] L. C. Schendel, M. S. Bauer, S. M. Sedlak, H. E. Gaub, *Small* **2020**, *16*, 1906740.

[79] S. Patra, J.-B. Claude, J.-V. Naubron, J. Wenger, *Nucleic Acids Res.* **2021**, *49*, 12348.

[80] M. P. Goldschen-Ohm, D. S. White, V. A. Klenchin, B. Chanda, R. H. Goldsmith, *Angew. Chem.* **2017**, *129*, 2439.

[81] J. Wenger, F. Conchonaud, J. Dintinger, L. Wawrezinieck, T. W. Ebbesen, H. Rigneault, D. Marguet, P.-F. Lenne, *Biophys. J.* **2007**, *92*, 913.

[82] C. V. Kelly, D. L. Wakefield, D. A. Holowka, H. G. Craighead, B. A. Baird, *ACS Nano* **2014**, *8*, 7392.

[83] J. M. Chandler, H. Xu, *AIP Adv.* **2021**, *11*, 065112.

[84] R. Regmi, P. M. Winkler, V. Flauraud, K. J. E. Borgman, C. Manzo, J. Brugger, H. Rigneault, J. Wenger, M. F. García-Parajo, *Nano Lett.* **2017**, *17*, 6295.

[85] P. M. Winkler, R. Regmi, V. Flauraud, J. Brugger, H. Rigneault, J. Wenger, M. F. García-Parajo, *J. Phys. Chem. Lett.* **2018**, *9*, 110.

[86] P. M. Winkler, R. Regmi, V. Flauraud, J. Brugger, H. Rigneault, J. Wenger, M. F. García-Parajo, *ACS Nano* **2017**, *11*, 7241.

[87] M. Baibakov, S. Patra, J.-B. Claude, A. Moreau, J. Lumeau, J. Wenger, *ACS Nano* **2019**, *13*, 8469.

[88] J. Widengren, P. Schwille, *J. Phys. Chem. A* **2000**, *104*, 6416.

[89] M. Levitus, S. Ranjit, *Q. Rev. Biophys.* **2011**, *44*, 123.

[90] C. Eggeling, A. Volkmer, C. A. M. Seidel, *ChemPhysChem* **2005**, *6*, 791.

[91] D. R. Lide, *CRC Handbook of Chemistry and Physics, 85th Edition*, CRC Press, **2004**.

[92] M. Fränzl, F. Cichos, *Nat. Commun.* **2022**, *13*, 656.

[93] Q. Jiang, B. Rogez, J.-B. Claude, G. Baffou, J. Wenger, *ACS Photonics* **2019**, *6*, 1763.

[94] Q. Jiang, B. Rogez, J.-B. Claude, A. Moreau, J. Lumeau, G. Baffou, J. Wenger, *Nanoscale* **2020**, *12*, 2524.

[95] K. Wang, E. Schonbrun, P. Steinvurzel, K. B. Crozier, *Nat. Commun.* **2011**, *2*, 469.

[96] M. Srisa-Art, E. C. Dyson, A. J. deMello, J. B. Edel, *Anal. Chem.* **2008**, *80*, 7063.

[97] J. Strömqvist, L. Nardo, O. Broekmans, J. Kohn, M. Lamperti, A. Santamato, M. Shalaby, G. Sharma, P. Di Trapani, M. Bondani, R. Rigler, *Eur. Phys. J. Spec. Top.* **2011**, *199*, 181.

[98] A. Tcherniak, C. Reznik, S. Link, C. F. Landes, *Anal. Chem.* **2009**, *81*, 746.

[99] U. Piran, W. J. Riordan, *J. Immunol. Methods* **1990**, *133*, 141.

[100] L. Deng, E. N. Kitova, J. S. Klassen, *J. Am. Soc. Mass Spectrom.* **2013**, *24*, 49.

[101] A. Chilkoti, P. S. Stayton, *J. Am. Chem. Soc.* **1995**, *117*, 10622.

[102] P. Roy, C. Badie, J.-B. Claude, A. Barulin, A. Moreau, J. Lumeau, M. Abbarchi, L. Santinacci, J. Wenger, *ACS Appl. Nano Mater.* **2021**, *4*, 7199.

[103] K. M. McPeak, S. V. Jayanti, S. J. P. Kress, S. Meyer, S. Iotti, A. Rossinelli, D. J. Norris, *ACS Photonics* **2015**, *2*, 326.

[104] J. Yang, J.-P. Hugonin, P. Lalanne, *ACS Photonics* **2016**, *3*, 395.

[105] S. Patra, M. Baibakov, J.-B. Claude, J. Wenger, *Sci. Rep.* **2020**, *10*, 5235.

[106] W. Schrimpf, A. Barth, J. Hendrix, D. C. Lamb, *Biophys. J.* **2018**, *114*, 1518.

**Supporting Information for**

**Achieving High Temporal Resolution in Single-Molecule Fluorescence Techniques using Plasmonic Nanoantennas**

Sunny Tiwari,[1] Prithu Roy,[1] Jean-Benoît Claude,[1] Jérôme Wenger[1,*]

[1] *Aix Marseille Univ, CNRS, Centrale Marseille, Institut Fresnel, AMUTech, 13013 Marseille, France*

*\* Corresponding author: jerome.wenger@fresnel.fr*

**Contents:**

S1. Comparison of the radiative rate constants of common fluorescent dyes
S2. Fabrication of the horn antenna using a focused ion beam
S3. Variation of count rate per molecule with respect to the microreflector cone angle
S4. Horn antenna does not change the near field enhancement in the aperture
S5. Redirecting maximum emission in narrow angles using horn antenna
S6. FCS time traces and correlation function in confocal, ZMW and horn antenna
S7. Variation of extracted number of molecules and diffusion time in horn antenna, ZMW, and confocal configuration with increasing input laser power
S8. Background measurements in the horn antenna and ZMW
S9. Statistical dispersion between different structures
S10. Larger collection gain using 0.65 NA Air objective lens
S11. Temperature measurement in the horn antenna and ZMW
S12. Dark state transition rate constants of Alexa Fluor 647 with temperature
S13. Enhanced fluorescence brightness of Alexa Fluor 546 dyes in horn antenna
S14. FCS data corresponding to the burst analysis
S15. High brightness and photon count per burst using horn antenna
S16. Visualizing single molecule bursts in ZMW at µs binning time
S17. Biotin-Streptavidin association rate calculation
S18. Additional measurement of biotin-streptavidin association rate for varied streptavidin concentration

**S1. Comparison of the radiative rate constants of common fluorescent dyes**

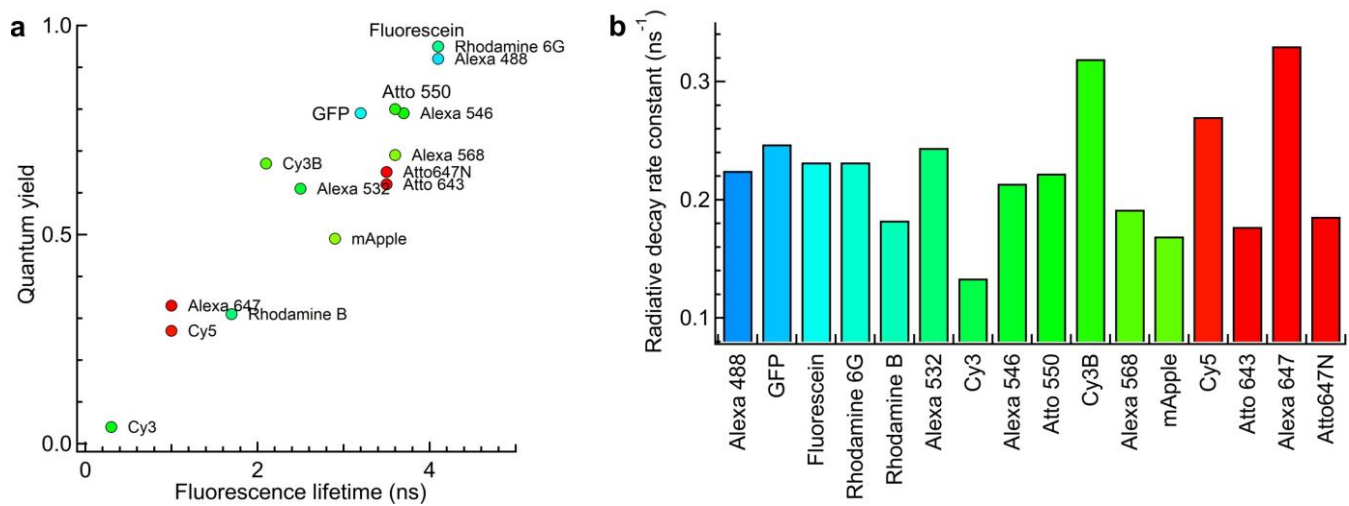

**Figure S1.** Alexa 647 is among the fluorescent dyes providing the highest radiative decay rate constant $k_R$. (a) Scatter plot of the quantum yield $\phi$ and the fluorescence lifetime $\tau$ for a selection of common fluorescent dyes. These values are used to compute the radiative rate constants $k_R = \phi * k_{TOT} = \phi/\tau$ which are displayed in (b).

**S2. Fabrication of the horn antenna using a focused ion beam**

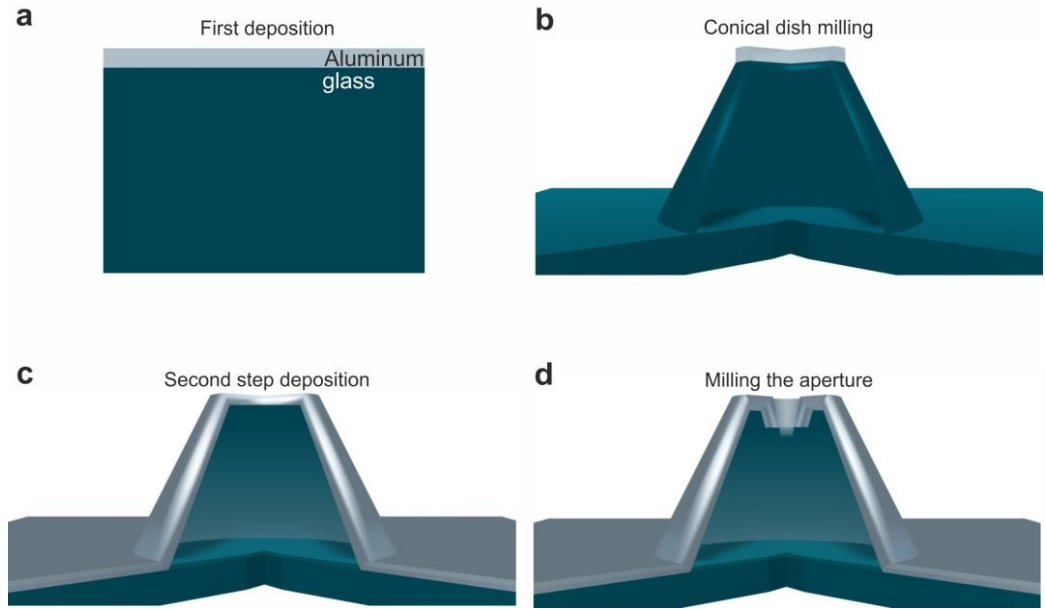

**Figure S2.** Step a) A glass coverslip of thickness 150 µm is plasma cleaned for use as a substrate. Over this glass substrate, a 100 nm aluminum layer is deposited which acts as an electrically conductive layer. Step b) Horn antenna with an upper plateau diameter of 1 µm and lower base diameter of 2-3 µm (depending on the angle of the reflecting arm) is milled by a focused ion beam (FIB) using a gallium ion source. Step c) Over this horn antenna, an additional 100 nm layer of aluminum layer is deposited to make the reflector wall surface reflective. Step d) a single nanohole of diameter 120 nm is milled in the centre of the upper plateau of the horn antenna. Finally, a 12 nm $SiO_2$ layer is also deposited on the top to protect the aluminum layer from corrosion and help with the surface passivation.

## S3. Variation of molecular brightness with respect to the microreflector cone angle

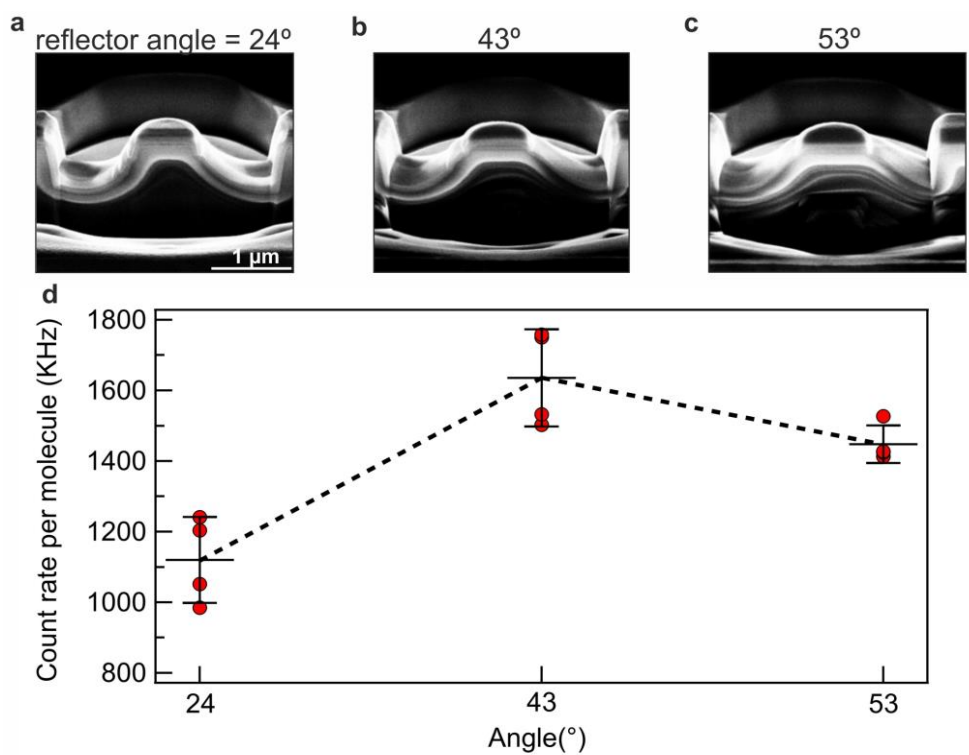

**Figure S3:** Variation of count rate per molecule with a change in the microreflector cone angle. (a-c) SEM images of horn antennas with arm angles of 24°, 43°, and 53°. (d) corresponding count rate per molecule obtained at an input power of 200 µW for Alexa 647 dyes at 60 nM concentration. Horn antennas with an arm angle of 43° were used in the experiments as the brightness decreases with either increasing or decreasing the arm angles of the horn antenna. For taking these representative SEM images, the sample is tilted by 52°. No apertures are milled in these representative SEM images. An complete characterization of the reflector angle influence was performed in Ref. [1] for the UV dye pterphenyl. The results shown here support the findings for the UV system and confirm that reflector angles around 40° ± 5° are near-optimum.

## S4. The horn antenna does not change the near field intensity enhancement inside the aperture

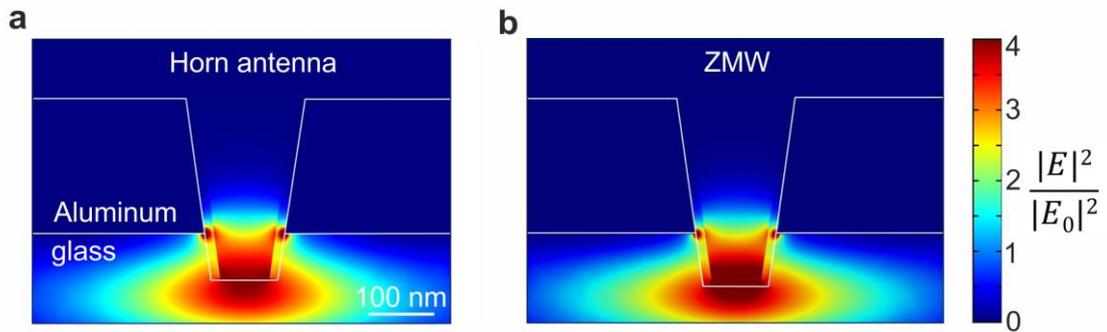

**Figure S4.** Electric field intensity enhancement in the horn antenna and the ZMW nanoaperture (without reflective conical structure). A similar value of the intensity enhancement is observed in the horn antenna (a) and in the ZMW (b) showing no change in the local near field enhancement because of the microreflector structure. The horn antenna and ZMW are excited from the bottom (glass) side using a Gaussian source of wavelength 633 nm.

## S5. Redirecting maximum emission in narrow angles using horn antennas

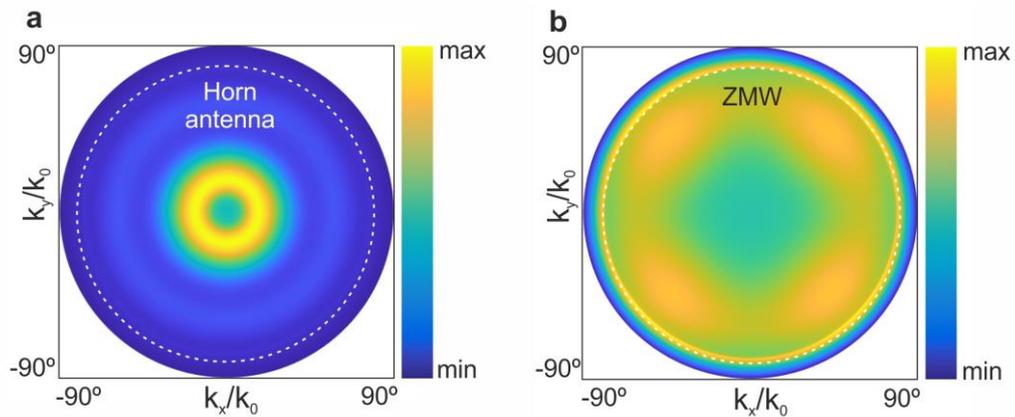

**Figure S5.** Calculated Fourier plane images of the emission from oscillating dipoles placed in the horn antenna (a) and ZMW (b). The emission from the horn antennas outcouples in a narrow range of angles as visible in the confinement in the calculated Fourier plane images. The Fourier plane images are calculated by incoherently adding the radiation pattern of the emission from each individual x, y, and z oriented dipoles placed in the center of the horn antenna and ZMW. The wavelength of the oscillations is set at 670 nm which is near the emission maxima of the Alexa 647 dyes used in the experiments. The white dotted circle represents the water-glass critical angle.

## S6. FCS time traces and correlation function in confocal, ZMW and horn antenna

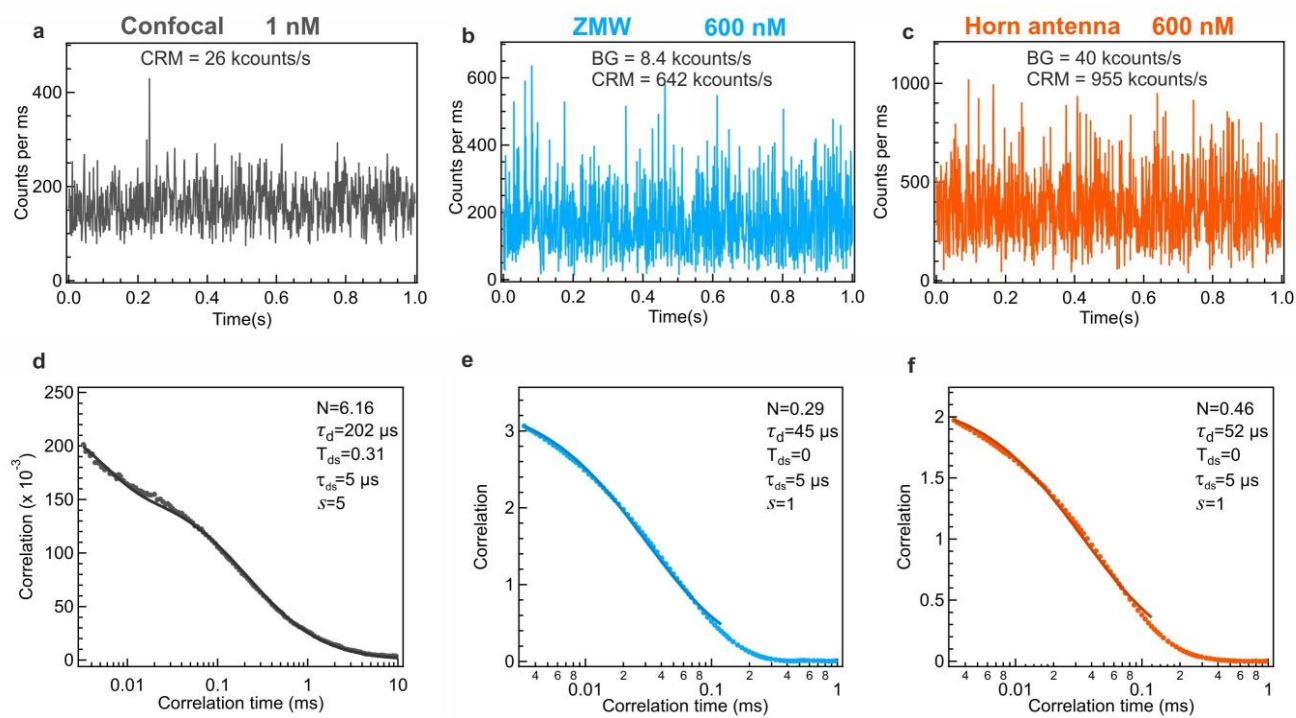

**Figure S6:** Fluorescence correlation spectroscopy of Alexa 647 molecules in the confocal configuration, ZMW, and horn antenna. Fluorescence intensity time trace (a-c) and corresponding FCS correlation function (d-f) for the confocal configuration, ZMW, and horn antenna respectively. The dotted curves are experimentally obtained FCS correlation amplitude and the solid curves are numerical fitting using the equation 1 for three dimensional Brownian motion with an additional blinking term as discussed in the method section. The concentration of Alexa 647 molecules is 1 nM for the confocal and 600 nM for horn antenna and ZMW. The input power of 633 nm laser is kept constant at 50 µW for all three configurations.

## S7. Variation of extracted number of molecules and diffusion time in horn antenna, ZMW, and confocal configuration with increasing input laser power

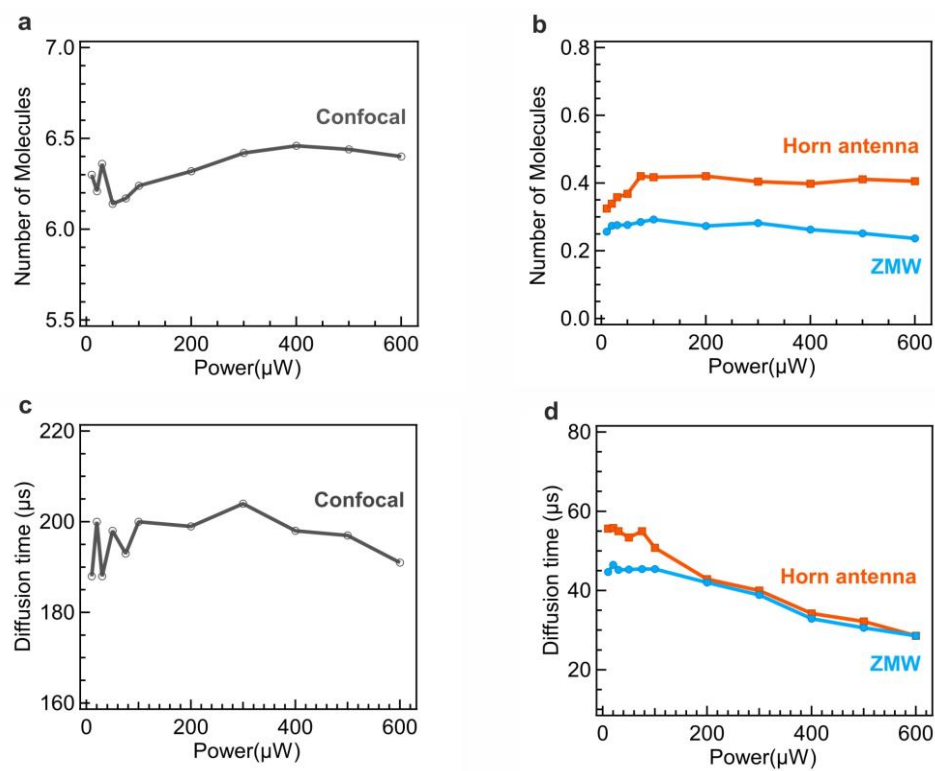

**Figure S7:** Variation in the extracted number of molecules (a, b) and diffusion time (c, d) as a function of input laser power for horn antenna, ZMW, and confocal configuration. Only a slight change in the number of molecules is observed for the confocal, horn antenna, and ZMW whereas the molecules diffuse much faster in the horn antennas and ZMW when the laser power is increased possibly due to the increment in the temperature of the surroundings which affects the water viscosity and induces thermal flows.

## S8. Background measurements in the horn antenna and ZMW

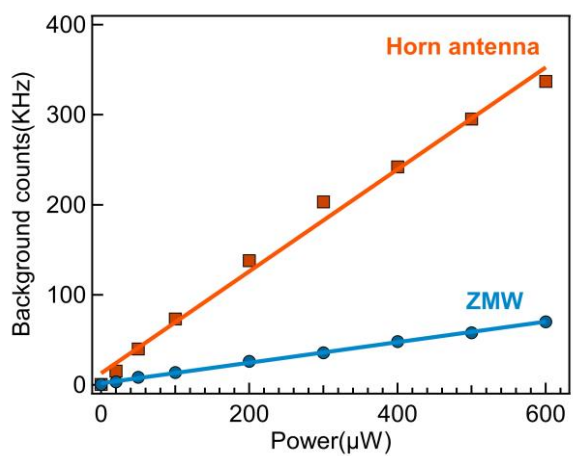

**Figure S8:** Background intensity from the horn antenna and ZMW as a function of the 633 nm laser power. A larger value of the background emission is obtained in the horn antenna as compared to the ZMW. This larger background is in part due to a better collection of the background luminescence from the substrate by the horn antenna. Another explanation for the higher background is related to the FIB milling of the microreflector, which involves more luminescent gallium oxide contamination.[2]

## S9. Statistical dispersion between different structures

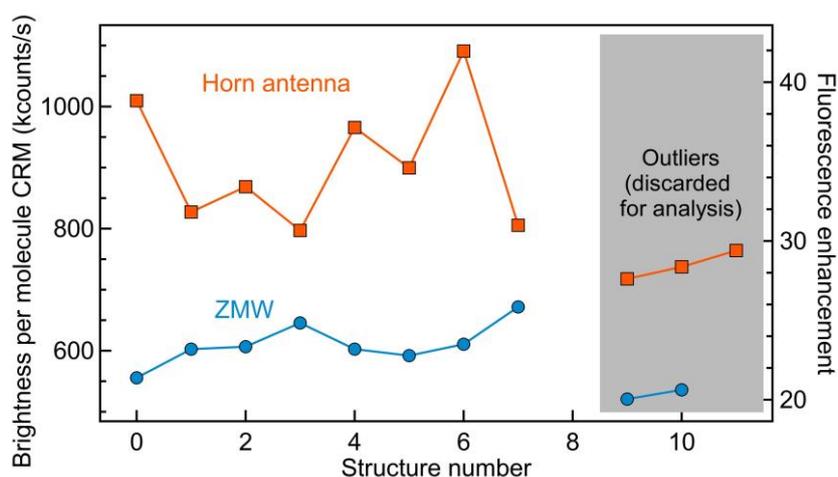

**Figure S9:** Fluorescence brightness per molecule measured for Alexa Fluor 647 in different horn antennas and ZMWs at 50 µW excitation power. For some structures (displayed in the shaded area), we found a brightness reduced by about 20% as compared to the average. For these structures, the FIB milling process may have been locally inefficient (or less efficient) due to uncontrolled variations of the parameters involved during the FIB milling (ion beam current, local beam distortions, local defects on the Al film, mechanical vibrations…). These structures have been discarded for the analysis.

## S10. Larger collection gain using 0.65 NA air objective lens

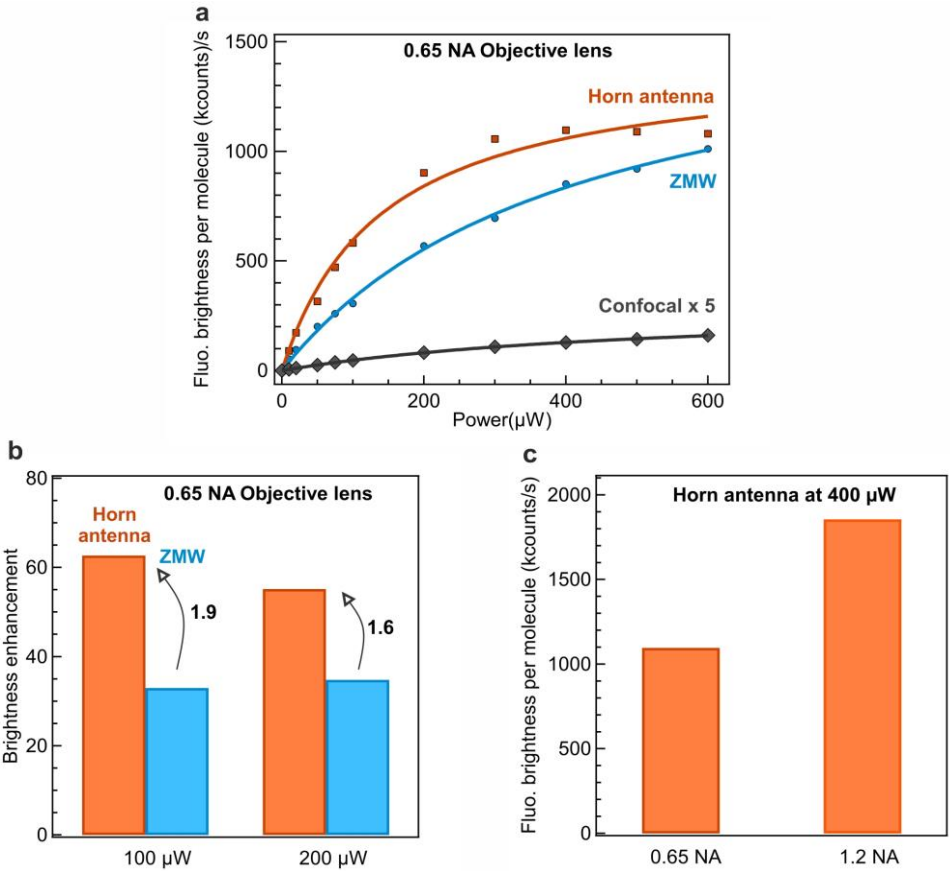

**Figure S10:** Larger collection gain using a lower NA (0.65 NA Air) objective lens. (a) Enhanced fluorescence brightness per molecule as a function of input power for Alexa 647 dyes in the horn antennas as compared to the ZMW and confocal reference using a 0.65 NA Air objective lens. A large enhancement can be achieved even by using low numerical aperture objective lenses, as the horn antenna redirects most of the emission in narrow angles, which can be collected efficiently even with a low NA lens. (b) Bar graph showing the brightness enhancement achieved in the horn antenna and ZMW using a 0.65 NA Air objective lens with respect to confocal reference at 100 and 200 µW. The concentration of Alexa 647 molecules in PBS buffer is 1 nM for confocal configuration and 600 nM for the horn antenna and ZMW.

## S11. Temperature measurement in the horn antenna and ZMW

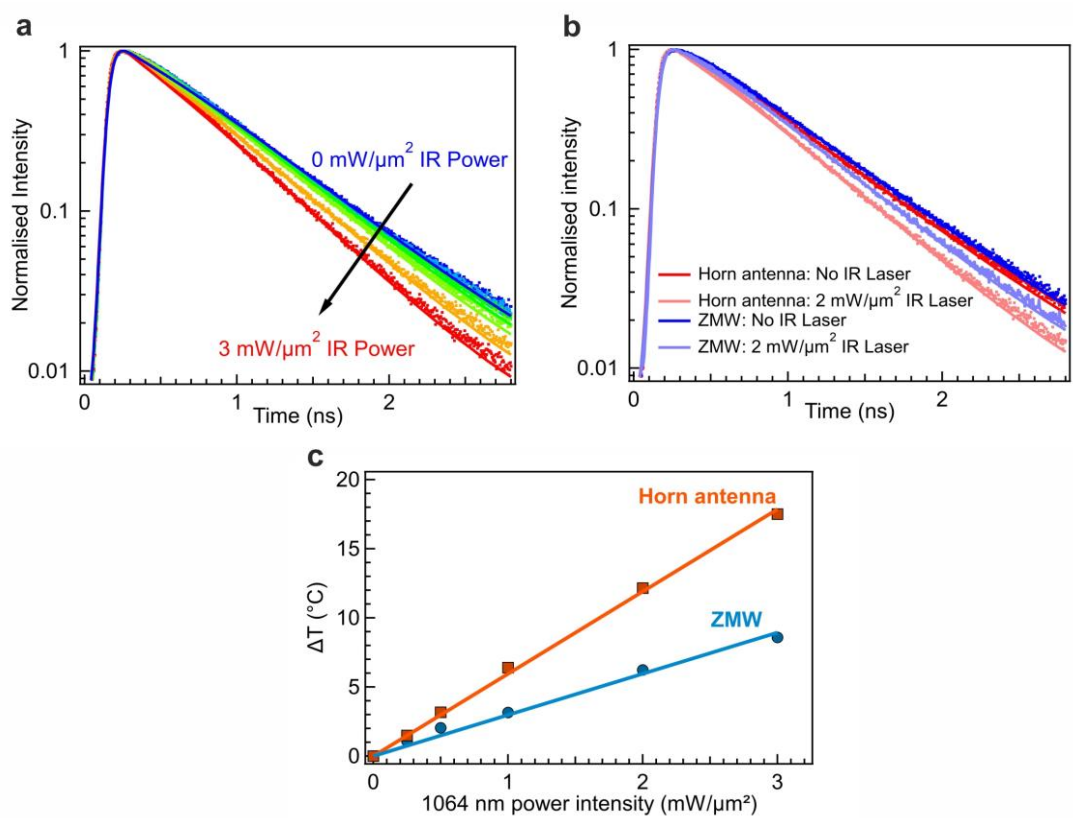

**Figure S11:** Temperature measurement in the horn antenna and ZMW. (a) Normalized fluorescence decay curves of Alexa 647 dyes in horn antennas at six different infrared power intensities (0, 0.25, 0.50, 1.0, 2.0, and 3.0 mW/μm$^2$: from top right to bottom left). (b) Normalized fluorescence decay curve of Alexa 647 dyes inside the horn antenna and ZMW without and with the infrared (1064 nm) laser excitation at 2 mW/μm$^2$. The Alexa 647 fluorescence lifetime decreases with the infrared laser illumination due to the higher local temperature, as previously discussed in Ref.[3,4]. (c) Estimated temperature increment for the horn antenna and ZMW as a function of 1064 nm laser power intensity following the procedure summarized below. The concentration of the Alexa 647 dyes used is 300 nM. The input power of 635 nm pulsed laser is kept constant at 15 μW/μm$^2$ where the power of infrared laser was varied from 0-3 mW/μm$^2$.

For the measurements of temperature change inside the optical horn antenna we used our recently developed technique to measure the temperature change in the ZMW by utilizing the variation of fluorescence lifetime of Alexa 647 dyes as a function of temperature change. Here we briefly discuss the fundamentals of this technique, a detailed methodology can be found in Ref.[3]. The fluorescence lifetime of the molecules is related to the radiative and non-radiative decay rates of the emission as following:

$$\frac{1}{\tau} = k_R + k_{NR}(T)$$

Where $k_R$ is the radiative decay rate constant, which is assumed to be independent of the temperature (verified for the temperature ranges studies in [3]). However, $k_{NR}$, the non-radiative decay rate constant, changes with the temperature variation, which further changes the lifetime of the molecules. This change in the fluorescence lifetime can be calibrated with a variation in the temperature of the surrounding. By putting the quantum yield of 33% at room temperature for Alexa 647 molecules, the change in the temperature ($\Delta T$) can be related to the change in the non-radiative decay rate ($k_{NR}$) as following:

$$\Delta T = 91.31 - 91.22 \times \left(\frac{k_{NR}(21^0 C)}{k_{NR}(T)}\right)^{0.42}$$

where $k_{NR}(21°C)$ is the non-radiative decay rate at room temperature. This temperature dependence has been calibrated in Ref.[3] and confirmed by two other methods based on fluorescence intensity and FCS diffusion time.

In the presence of the horn antenna, the lifetime of the Alexa dyes gets modified due of the change in the radiative decay rate because of the Purcell enhancement and introduction of the new decay channels due to the presence of the metals nearby. Upon laser excitation of the optical horn antenna, the non-radiative decay rates also get modified because of the change in the temperature of the surrounding due to the laser heating of the horn antenna. In addition with the radiative decay rate, the losses due to the coupling of the emission to the metal also does not depend on the temperature change. Thus, equation 1 becomes

$$\frac{1}{\tau_{T^*}} = k_R^* + k_{loss}^* + k_{NR}^*(T)$$

where $k^*_{loss}$ represents the quenching of the emission because of the coupling of the emission to the metal. Increasing the temperature of the surrounding of horn antenna results in a shortening of the lifetime of Alexa 647 fluorescence emission and thus the normalized decay rate inside the optical horn antenna becomes;

$$\frac{k_{NR}(T)}{k_{NR}(21^0 C)} = 1 + \frac{1}{k_{NR}(21^0 C)}\left(\frac{1}{\tau_{T^*}} - \frac{1}{\tau_{21^0 C^*}}\right)$$

The quantum yield of the Alexa 647 molecules used has a quantum yield of 33% with a lifetime of 1 ns which give a value 0.67 ns$^{-1}$ for $k_{NR}(21°C)$. Finally, we get the equation to calculate the temperature change around the optical horn antenna as following:

$$\Delta T = 91.31 - 91.22 \times \left[1 + \frac{1}{0.67}\left(\frac{1}{\tau_{T^*}} - \frac{1}{\tau_{21^0C^*}}\right)\right]^{-0.42}$$

To extract the temperature change in the horn antenna for 633 nm laser excitation, we first extract the change in the temperature of horn antenna as a function of a 1064 nm infrared laser. In the horn antenna, the absorbance of light by the aluminum metal is the main source of heating effects. We utilize the ratio of absorbance of aluminum at 1064 nm and 633 nm laser to extrapolate the temperature increment in the horn antenna at 633 nm excitation. The complex refractive index of aluminum at 633 nm and 1064 nm are taken from Ref.[5,6]

All the fluorescence decay histograms are fitted using a Levenberg-Marquard optimization using *Picoquant SymPhoTime 64* software. Iterative re-convolution based fitting is performed by taking into account the instrument response function. We use a bi-exponential function to fit the decay curves where the shorter lifetime corresponds to the losses because of the coupling to the metal and the larger lifetime parameter corresponds to the fluorescence lifetime change related to the temperature variation. The temperature measurements are performed on an inverted microscope equipped with two overlapping beams of wavelength 635 nm pulsed laser (*Picoquant* LDH-P-635, 80 MHz repetition rate) and 1064 nm infrared laser (*Ventus* 1064-2W). A high NA oil immersion objective lens (40x, NA 1.3, *Zeiss Plan-Neofluar*) is used to focus both the lasers on the sample. The measured spot size for the 635 nm and 1064 nm lasers are 0.6 µm and 1.0 µm respectively. A low power 635 nm pulsed laser at 15 µW was used to excite the Alexa 647 fluorescence, to simultaneously avoid, the heating of the horn antenna by the red laser, and the photobleaching and saturation of the Alexa molecules. In addition, the diffusing molecules in the solution also acts as a fresh source of fluorescent molecules, thus further reducing the chances of photobleaching and saturation effects. The power of the infrared laser was tuned from 0 to 3 mW/µm² to heat the horn antenna and ZMW. No sign of damage to either the horn antenna or zero mode waveguide was observed (confirmed by the measurement of fluorescence signal before and after the IR laser exposure).

A 300 nM concentration of Alexa 647 molecules in phosphate buffered saline solution was used to probe the fluorescence and lifetime variation as a function of infrared laser power. The fluorescence of the molecules was excited in the epifluorescence configuration. The emission was spatially filtered using a 50 µm pinhole and was directed to two avalanche photodiodes (*Picoquant MPD-5CTC*) to record the emission. Next, the signal was collected by a time-correlated single photon counting system (*Picoquant Picoharp300*) with time-tagged time-resolved (TTTR) option, with an overall temporal resolution (full width at half maximum of the instrument response function) of 100 ps.

## S12. Dark state transition rate constants of Alexa Fluor 647 with temperature

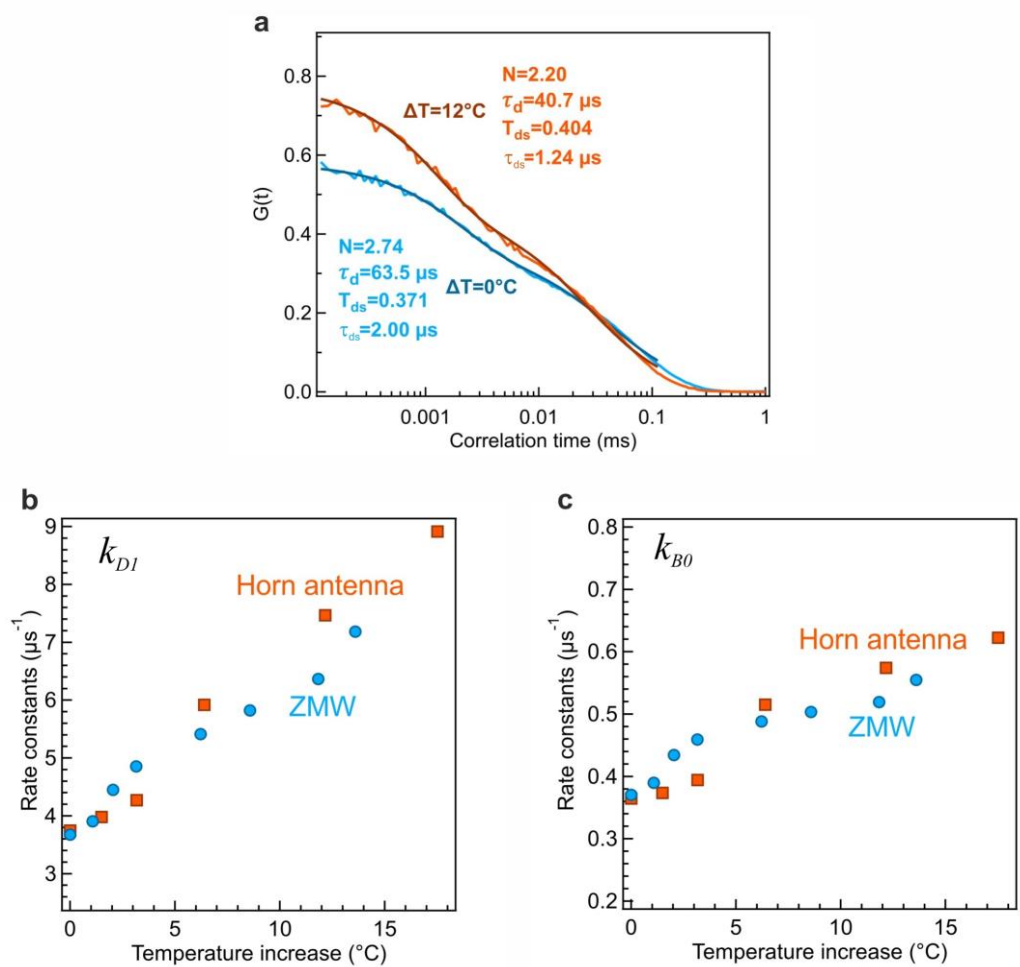

**Figure S12:** Variation of the transition rate constants $k_{D1}$ and $k_{B0}$ to and from the dark state for Alexa Fluor 647 dye as a function of temperature increase in the horn antenna and ZMW. (a) FCS correlation curves and numerical fitting of Alexa 647 dyes in horn antenna at room temperature and at temperature increment of 12°C (set by an additional 1064 nm laser as used in section S11). The use of two APDs along with the Fluorescence Lifetime Correlation Spectroscopy (FLCS) background correction removes the effects of afterpulsing, enabling to retrieve the accurate information on microsecond dynamics. The extra infrared laser allows to accurately control the local temperature and simultaneously measure the temperature influence on $k_{D1}$ and $k_{B0}$. (b, c) Evolution of the dark state transition $k_{D1}$ and $k_{B0}$ rate constants as a function of the temperature increment deduced from FCS measurements in the horn antenna and ZMW.

To determine experimentally $k_{D1}$ and $k_{B0}$, we use the FCS fit results $T_{ds}$ and $\tau_{ds}$ of the fast blinking term in the correlation function (Eq. (3) in the main text). The experiments in Fig. S11 aims at measuring the dependence of $k_{D1}$ and $k_{B0}$ with the temperature. To this end, we have used a supplementary infrared laser to control the local temperature (as used in the experiments of Fig. S11 and described in[3]). The 15 µW power at 635 nm excitation of Alexa 647 remains in the linear regime well below saturation to clearly discriminate the effect of the local temperature. In this condition, the FCS quantities $T_{ds}$ and $\tau_{ds}$ are related to the decay rate constants by:[7,8]

$$T_{ds} = \frac{k_{D1}}{k_{B0}} \frac{k_{EX}}{k_R + k_{NR}}$$

$$\frac{1}{\tau_{ds}} = k_{B0} + k_{D1} \frac{k_{EX}}{k_R + k_{NR}}$$

This set of equations can be solved to yield $k_{D1}$ and $k_{B0}$ as a function of measurable quantities:

$$k_{D1} = \frac{T_{ds}}{1 + T_{ds}} \frac{k_R + k_{NR}}{k_{EX}} \frac{1}{\tau_{ds}}$$

$$k_{B0} = \frac{1}{1 + T_{ds}} \frac{1}{\tau_{ds}}$$

For the excitation rate, we have used the 15 µW average power, which is equivalent to a photon intensity of 1.45e22 photons/s/cm². Together with the absorption cross section $\sigma$ = 1e-15 cm² of Alexa 647, and a 4-fold local intensity enhancement (Fig. S4), we obtain an excitation rate $k_{EX}$ of 0.06 ns⁻¹. The sum $k_R + k_{NR}$ is equivalent to the total decay rate constant in the nanoaperture, which is taken as the inverse of the 600 ps fluorescence lifetime for the experiments in Fig. S12.

## S13. Enhanced fluorescence brightness of Alexa 546 dyes in horn antenna

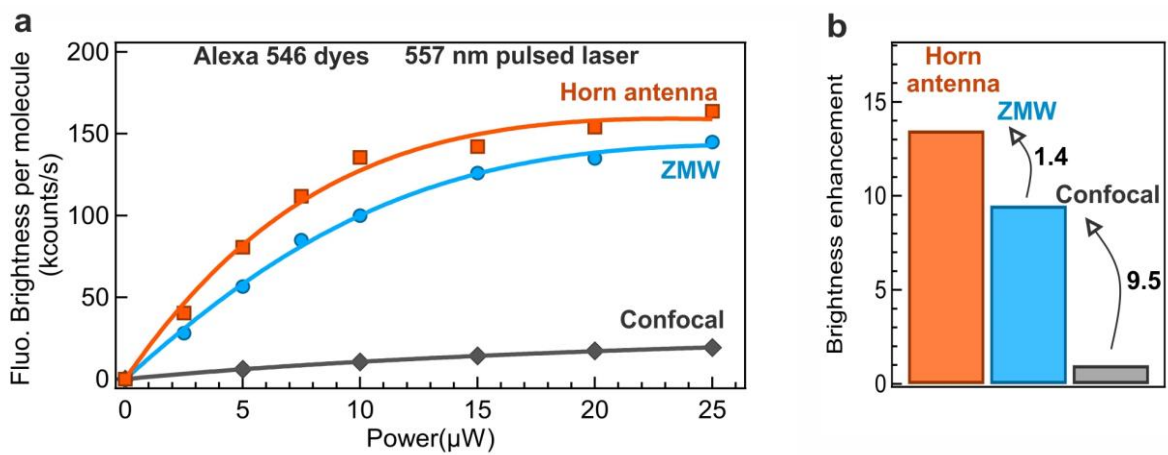

**Figure S13:** Enhanced fluorescence brightness per molecule of Alexa Fluor 546 dyes in the horn antenna. (a) Variation of fluorescence brightness per molecules in the horn antenna, ZMW, and confocal configuration as a function of input laser power of wavelength 557 nm. (b) Bar plot showing a larger enhancement of the fluorescence brightness per molecule in the horn antenna as compared to the ZMW and confocal configuration for a laser power of 10 µW. The concentration of the Alexa 546 dyes in horn antenna and ZMW is 200 nM and in confocal configuration is 1 nM. Full details about the experimental setup are given in Ref.[9] Here the Alexa 546 dyes are labeling a 51 base pair double stranded DNA molecule.

## S14. FCS data corresponding to the burst analysis

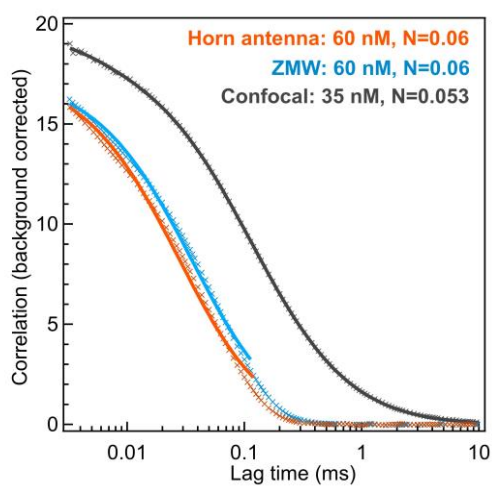

**Figure S14:** Background corrected FCS amplitude $G/\left(1-\frac{B}{F}\right)^2$ of Alexa 647 dyes in the horn antenna, ZMW, and confocal configuration. The extracted number of molecules are ~0.05 for confocal, and 0.06 for horn antenna and ZMW. The concentration of the Alexa 647 dyes used is 60 nM in the horn antenna and 35 pM in the confocal reference. The correlation function is normalized with respect to the background in all three configurations resulting in approximately an equal number of molecules. The input power of 633 nm laser is 200 μW.

## S15. High brightness and photon counts per burst using horn antenna

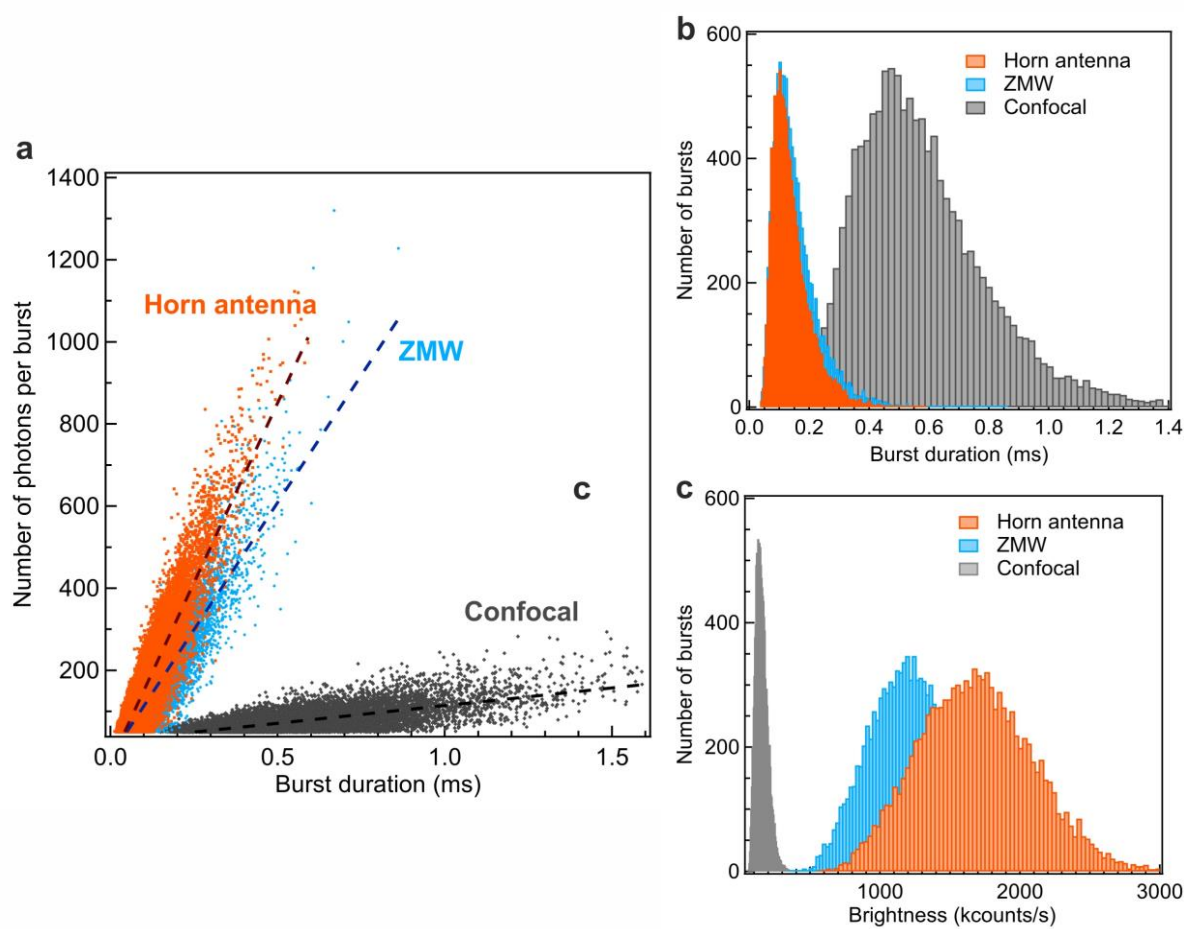

**Figure S15:** (a) Scatter plots of the integrated intensity per burst as a function of the burst duration, comparing the horn antenna, the ZMW and the confocal reference. Each dot represents a detected fluorescence burst. In the horn antenna and ZMW, the bursts are of shorter duration as a consequence of the volume reduction as compared to the diffraction limited confocal configuration. (b) Histograms of the burst durations showing a larger number of short duration bursts in the horn antenna and ZMW compared to the confocal reference where the bursts are of longer duration. (c) Histograms of the fluorescence brightness for each burst, showing a larger number of bursts with very high brightness (reaching 2 million counts/s) in the horn antennas overpassing the brightness achieved in the ZMW and confocal configuration.

**S16. Visualizing single molecule bursts in ZMW at microsecond binning time**

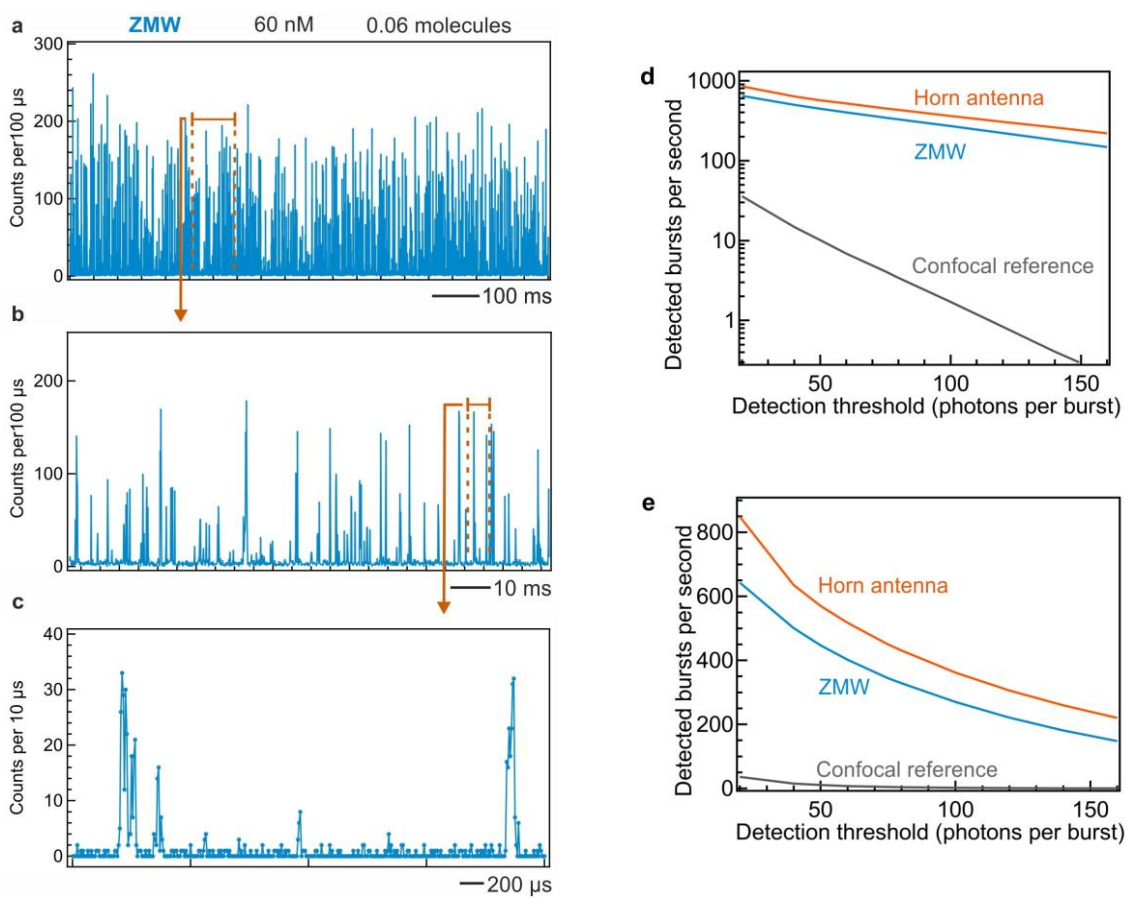

**Figure S16:** (a) Fluorescence time trace of Alexa 647 dyes in the ZMW with 100 μs binning time. (b) Zoomed in time trace of the dotted rectangular region in (a). (c) Time trace at 10 μs bin time of the rectangular region shown in (b) showing the well resolved single molecular bursts. The concentration of the Alexa 647 dyes in PBS solution used is 60 nM corresponding to 0.06 average number of molecules in the ZMW. The input power of 633 nm laser used is 200 μW. (d,e) Number of detected bursts per second as a function of the detection threshold, adding the ZMW data on Fig. 3h to allow an easy comparison between the horn antenna, ZMW and confocal reference. Both graphs (d,e) display the same data, (d) has a logarithmic vertical scale while (e) is in linear scale. The horn antenna improves the ZMW values by approximately 30%.

**S17. Biotin-Streptavidin association rate calculation**

We study the binding and unbinding reaction dynamics of the form $S + B \rightleftarrows SB$ where $S$ denotes the free streptavidin, $B$ the free biotin and $SB$ is the complex. In our case, the biotin concentration is the limiting factor and is always significantly lower than the streptavidin concentration. Therefore, we may safely assume that only a single biotin binds to a streptavidin. This way we may not take into account the fact that there are 4 binding sites for biotin on a single streptavidin. Cooperativity is not taken into account here for the same reasons.

The association rate constant to form the complex is noted $k_1$ while the dissociation rate constant is $k_{-1}$. The time origin t=0 is set when streptavidin is loaded into the biotin solution. One can write a set of differential equations to study the reaction kinetics:

$$\frac{d[B]}{dt} = -k_1[S][B] + k_{-1}([B_{tot}] - [B])$$

$$\frac{d[SB]}{dt} = k_1[S][B_{tot}] - (k_1[S] + k_{-1})[SB]$$

Where the total biotin concentration $[B_{tot}] = [B](t) + [SB](t)$ is a constant. The set of differential equations can be solved with the boundary condition that $[B](0) = [B_{tot}]$ and $[SB](0) = 0$. The free biotin fraction is then:

$$\frac{[B](t)}{[B_{tot}]} = e^{-(k_1[S]+k_{-1})t} + \frac{k_{-1}}{k_1[S] + k_{-1}}\left(1 - e^{-(k_1[S]+k_{-1})t}\right)$$

The second term in the right hand side is a small offset correction which can be easily neglected if the dissociation $k_{-1}$ is much lower than the association $k_1[S]$, so we have simply an exponential decay.

The bound biotin fraction is:

$$\frac{[SB](t)}{[B_{tot}]} = \frac{k_1[S]}{k_1[S] + k_{-1}}\left(1 - e^{-(k_1[S]+k_{-1})t}\right)$$

Both the free and the bound biotin fraction follow an exponential temporal evolution $e^{-t/\tau}$ with a characteristic time $\tau = 1/(k_1[S] + k_{-1})$. Experimentally we determine by FCS the correlation amplitudes $\rho_1$ and $\rho_2$ corresponding to the free and bound biotin respectively based on the difference in the diffusion times. The ratios give the free and bound biotin fractions:

$$\frac{\rho_1}{\rho_1+\rho_2} = \frac{[B](t)}{[B_{tot}]} \quad \text{and} \quad \frac{\rho_2}{\rho_1+\rho_2} = \frac{[SB](t)}{[B_{tot}]}$$

Fitting these ratios with an exponential function gives the rate constant $k_1[S] + k_{-1}$. The dissociation rate constant $k_{-1}$ is 2e-6 s$^{-1}$ at 20°C for streptavidin-biotin,[10] so we can easily neglect this term.

## S18. Additional measurements of biotin-streptavidin association for different streptavidin concentrations

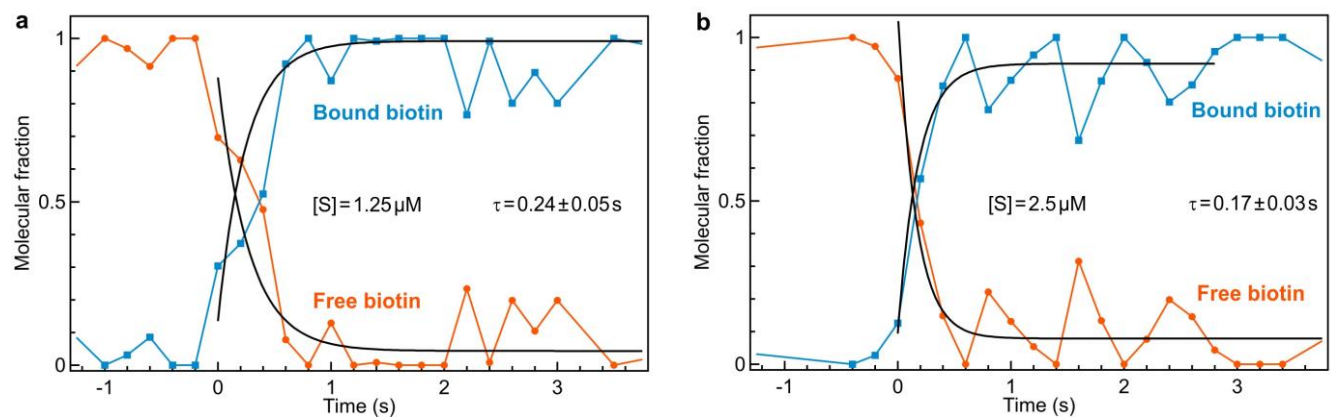

**Figure S17:** Characteristic time variation as a function of streptavidin concentration used. A characteristic time of 0.24 s (0.17 s) is obtained for a streptavidin concentration of 1.25 µM (2.5 µM) for a fixed Biotin-Atto643 concentration of 0.5 µM. The integration time of the FCS measurement is 200 ms. The aperture diameter of 200 nm is used at 50 µW input power of 633 nm laser. All the FCS curves are fitted using a two species model,[11] where the FCS diffusion times of each species (free and bound) are fixed according to the values found for 30 s integration time in Fig. 4a. The evolution of the molecular fraction of bound and free biotins are performed following the exponential evolution derived in section S17.

**Supplementary references**

[1] A. Barulin, P. Roy, J.-B. Claude, J. Wenger, *Nat. Commun.* **2022**, *13*, 1842.

[2] P. Roy, J.-B. Claude, S. Tiwari, A. Barulin, J. Wenger, *Nano Lett.* **2023**, *23*, 497–504.

[3] Q. Jiang, B. Rogez, J.-B. Claude, G. Baffou, J. Wenger, *ACS Photonics* **2019**, *6*, 1763.

[4] Q. Jiang, B. Rogez, J.-B. Claude, A. Moreau, J. Lumeau, G. Baffou, J. Wenger, *Nanoscale* **2020**, *12*, 2524.

[5] K. M. McPeak, S. V. Jayanti, S. J. P. Kress, S. Meyer, S. Iotti, A. Rossinelli, D. J. Norris, *ACS Photonics* **2015**, *2*, 326.

[6] M. N. Polyanskiy, Refractive index database, https://refractiveindex.info, accessed: Jul., 2019.

[7] J. Widengren, R. Rigler, Ü. Mets, *J. Fluoresc.* **1994**, *4*, 255.

[8] J. Wenger, B. Cluzel, J. Dintinger, N. Bonod, A.-L. Fehrembach, E. Popov, P.-F. Lenne, T. W. Ebbesen, H. Rigneault, *J. Phys. Chem. C* **2007**, *111*, 11469.

[9] M. Baibakov, S. Patra, J.-B. Claude, A. Moreau, J. Lumeau, J. Wenger, *ACS Nano* **2019**, *13*, 8469.

[10] L. Deng, E. N. Kitova, J. S. Klassen, *J. Am. Soc. Mass Spectrom.* **2013**, *24*, 49.

[11] J. Strömqvist, L. Nardo, O. Broekmans, J. Kohn, M. Lamperti, A. Santamato, M. Shalaby, G. Sharma, P. Di Trapani, M. Bondani, R. Rigler, *Eur. Phys. J. Spec. Top.* **2011**, *199*, 181.